\documentclass[fleqn,usenatbib]{mnras}

\usepackage[T1]{fontenc}

\DeclareRobustCommand{\VAN}[3]{#2}
\let\VANthebibliography\thebibliography
\def\thebibliography{\DeclareRobustCommand{\VAN}[3]{##3}\VANthebibliography}

\usepackage{graphicx}	
\usepackage{amsmath}	
\usepackage{amssymb}	

\usepackage{newtxtext,newtxmath}

\usepackage{listings}

\newcommand{\ocen}{\ensuremath{\omega\,\mathrm{Cen}}}

\title[Tidal debris from $\ocen$]{Tidal debris from Omega Centauri discovered with unsupervised machine learning}

\author[K. Youakim et al.]{
Kris Youakim$^{1}$\thanks{E-mail: kristopher.youakim@astro.su.se},
Karin Lind$^{1}$,
and Iryna Kushniruk$^{1}$
\\
$^{1}$Department of Astronomy, Stockholm University, AlbaNova University Centre, Roslagstullsbacken 21, 106 91 Stockholm, Sweden\\
}

\date{Accepted 2023 June 19. Received 2023 June 15; in original form 2022 November 9}

\pubyear{2023}

\begin{document}
\label{firstpage}
\pagerange{\pageref{firstpage}--\pageref{lastpage}}
\maketitle

\begin{abstract}
The gravitational interactions between the Milky Way and in-falling satellites offer a wealth of information about the formation and evolution of our Galaxy. In this paper, we explore the high-dimensionality of the GALAH DR3 plus Gaia eDR3 data set to identify new tidally stripped candidate stars of the nearby star cluster Omega Centauri ($\ocen$). We investigate both the chemical and dynamical parameter space simultaneously, and identify cluster candidates that are spatially separated from the main cluster body, in regions where contamination by halo field stars is high. Most notably, we find candidates for $\ocen$ scattered in the halo extending to more than $50^{\circ}$ away from the main body of the cluster. Using a grid of simulated stellar streams generated with $\ocen$ like orbital properties, we then compare the on sky distribution of these candidates to the models. The results suggest that if $\ocen$ had a similar initial mass as its present day mass, then we can place a lower limit on its time of accretion at t$_{\mathrm{acc}} > 7$ Gyr ago. Alternatively, if the initial stellar mass was significantly larger, as would be expected if $\ocen$ is the remnant core of a dwarf Galaxy, then we can constrain the accretion time to t$_{\mathrm{acc}} > 4$ Gyr ago. Taken together, these results are consistent with the scenario that $\ocen$ is the remnant core of a disrupted dwarf galaxy.


\end{abstract}




\section{Introduction}

In the paradigm of $\Lambda$CDM cosmology, substructures in the halo of the Milky Way are the remnants of past accretion events from smaller stellar systems that are pulled into the much larger gravitational potential of the Milky Way. As these systems orbit our Galaxy, their stars are stripped through tidal interactions and are dispersed throughout the halo. During this process these stars retain a signature of their birthplace and can be distinctly identified from other Milky Way stars as co-moving groups with similar chemical abundances \citep[e.g.][]{Eggen1970, Freeman_Hawthorn2002}.  

Therefore, the present day locations, kinematics, and chemistry of accreted Milky Way stars can help to associate them to their original progenitor \citep[e.g.][]{Eggen1970, Myeong2018b, Myeong2022}. Indeed, numerical simulations have also demonstrated that stars stripped from their parent cluster can still be identified as coherent groups in phase-space, even over long periods of cosmic time \citep[e.g.][]{Helmi1999, Meza2005}. Likewise, coeval stars with similar chemical abundance ratios can be traced back to their common birth cluster through a process referred to as chemical tagging \citep[e.g.][]{Freeman_Hawthorn2002, Bland-Hawthorn2010}. Several recent works have investigated the feasibility of chemical tagging in our Galaxy \citep[e.g.][and references therein]{Ting2015, Quillen2015, Price-Jones_Bovy2019, Hawkins2020}, demonstrating some promising preliminary results \citep{Hogg2016, Kos2018, Price-Jones2020}, but also some limitations \citep{Mitschang2014, Blanco-Cuaresma2015}. 

Typically, star clusters are identified by the spatial clustering of their member stars. Using clever signal processing techniques, these overdensities can be found even when the signal is faint \citep[e.g.][]{Belokurov2006, Krone-Martins2014, Pera2021}. However, finding stars far away from the main cluster body is challenging due to the overwhelming abundance of foreground and background field stars, although this is sometimes possible with deep imaging and very good data \citep{Carballo_Bello2018, Kuzma2018}. When available, additional information such as chemistry and kinematics can be used to associate stars to their parent cluster, even when they are scattered in the halo and do not form a coherent stream \citep[e.g.][]{Majewski2012, Lind2015, Simpson2020}.



The nearby star cluster Omega Centauri ($\ocen$) has been extensively studied and shown to have many peculiarities in its stellar populations. First, a multipeaked metallicity distribution function covering a large range $-0.5 \gtrsim \textrm{[Fe/H} \gtrsim -2$ \citep{Suntzeff1996, Lee1999, Pancino2000, Johnson2010}, with these metallicity peaks corresponding to multiple sequences for the red giant branch \citep{Lee1999, Pancino2000, Sollima2005}, sub-giant branch \citep{Ferraro2004, Villanova2007, Villanova2014}, and main sequence \citep{Anderson1997, Bedin2004, Latour2021}. In addition, several works have described enhancements, and large spreads in abundances of light elements such as [Na/Fe], [O/Fe], and [Al/Fe] of > 0.5 dex, as well as strong anticorrelations in Na--O and Al--O  \citep{Brown_Wallerstein1993, Norris_DaCosta1995a, Johnson2010, Carretta2010, Marino2011}, and more recently an anticorrelation in Mg--K as well \citep{Alvarez2022}. Furthermore, $\ocen$ stars have been shown to be enriched in [Ba/Fe] compared to halo stars of similar metallicities, with a knee at [Fe/H] < $-1.5$ \citep{Norris_DaCosta1995b, Smith2000, Majewski2012}, and have an extreme overabundance of [Ba/Eu] $\sim$ 1 at [Fe/H] $\sim$ -1 \citep[see for example figure 11 in][]{Geisler2007}. Furthermore \citet{Smith2000} show that the [Cu/Fe] abundance in $\ocen$ stars is low, at [Cu/Fe] = $-0.6$, and remains constant with increasing metallicity (see their figure 9), whereas in field stars, [Cu/Fe] is expected to increase with metallicity \citep[e.g.][]{Sneden_Gratton1991}. 


The dynamical properties of $\ocen$ have also been thoroughly investigated. It has an orbital period of $\sim$ 120 Myr, characterized by low orbital energy and inclination, and is distinctly retrograde ($L_z$ < 0) in comparison to other clusters with similar metallicity and orbital energy \citep{Dinescu1999}. Furthermore, it has also been suggested that $\ocen$ is the remnant core of a dwarf galaxy \citep[e.g.][]{Bekki2003, Carretta2010}, which would imply that there should be some tidal debris dispersed in the halo from this disruption event. Indeed, previous works have identified extended, retrograde material in the Galactic halo that the authors have suggested is likely to have been stripped from $\ocen$ \citep{Dinescu2002, Majewski2012, Myeong2018b}. In another study, \citet{Ibata2019b} showed a connection between the main body of $\ocen$ and the recently discovered Fimbulthul stellar stream \citep{Ibata2019a}, using the orbital properties of their stars along with an $N$-body simulation. This finding was later corroborated with the discovery of two stars linking Fimbulthul and $\ocen$ using chemical tagging \citep{Simpson2020}. However, the stream stars could not be traced all the way to the body of $\ocen$ due to significant crowding at low Galactic latitudes. 

The Galactic archaeology with Hermes \citep[GALAH;][]{De_Silva2015} survey has observed hundreds of thousands of stars with high resolution spectroscopy and has recently published its third public data release \citep[GALAH DR3;][]{Buder2021}. The detailed analysis of these spectra has provided abundances for up to 30 chemical elements, and when combined with the early third data release from the Gaia mission \citep[Gaia eDR3:][]{Gaia2016, Gaia2021, Lindegren2021}, it also provides several dynamical and orbital parameters. Taken together, this is a powerful high-dimensional data set with which to search for cluster members that have been ejected from their parent cluster through dynamical interactions with the Milky Way. 



A recent study by \citet{Kos2018} used GALAH DR1 \citep{Martell2017} data to perform a search for additional members of nine open and globular clusters contained in the GALAH survey footprint. They selected stars within a radius around each cluster, and using the measured abundances of 13 elements they applied a dimensionality reduction algorithm to chemically tag cluster members. They were able to recover the known members of seven out of the nine clusters in the 2D latent space and demonstrate the feasibility of this method. They also found two new members of the Pleiades open cluster, which were confirmed to have been ejected based on their kinematic properties, but did not claim any new members for any of the other clusters. In this paper, we expand upon the work of \citet{Kos2018}, and add orbital parameters as input into the dimensionality reduction, implementing simultaneous chemical and kinematic tagging to identify tidally stripped stars from Omega Centauri. We also use an updated data set with more stars and much improved abundances and kinematics, namely GALAH DR3 and Gaia eDR3.

In Section \ref{sec:methods} of this paper we describe the methodology used to identify cluster members, Section \ref{sec:results} presents the new candidates including an analysis of their kinematics and chemistry, and in Sections \ref{sec:discussion} and \ref{sec:conclusions} we compare to mock stream models and discuss the implications of these newly discovered candidates for the accretion scenario of $\ocen$ and provide our conclusions from this work.



\section{Methods}
\label{sec:methods}

In this work, we utilized an unsupervised clustering algorithm called t-distributed stochastic neighbour embedding (t-SNE). In short, this method implements non-linear dimensionality reduction through initialization of N-dimensional probability distributions and minimization of a loss function to efficiently map clustering in high-dimensional space to lower dimensions. For a detailed description of the methodology (upon which this paper expands) see \citet{Kos2018}, and for more details of the technical aspects of t-SNE and comparisons to other forms of dimensionality reduction, we refer the reader to \citet{Hinton2002} and \citet{vanderMaaten2008}. We also experimented with other algorithms for dimensionality reduction, such as Uniform Manifold Approximation and Projection for Dimension Reduction \citep[UMAP;][]{McInnes2018} and a Variational Auto-encoder \citep[VAE;][]{Kingma_Welling2013}, but found that t-SNE performed the best empirically in terms of recovering the known literature members of $\ocen$ in the latent space projection.

We implemented the MulticoreTSNE python package \citep{Ulyanov2016}, which is a multicore modification of Barnes-Hut t-SNE \citep{vanderMaaten2013}, an efficient and fast implementation with python and Torch CFFI-based wrappers. For input data with 20 parameters and $\sim 50,000$ stars (the number of stars in the $\ocen$ sample), one run took about 4 min on an ordinary eight core laptop. This meant that the full analysis for a given set of input parameters of 100 bootstraps could be run in about 6--7 h.

\subsection{Sample selection}
\label{sample_selection}

The abundances and kinematic parameters used for the stars in our sample were taken from the latest data release from the GALAH survey \citep[DR3;][]{Buder2021}. We cross-matched the GALAH\_DR3\_main\_allstar\_v2.fits table with the value added catalogues (VACs) provided as supplements to the third data release. The stellar parameters and abundances are from the main catalogue, GALAH\_DR3\_main\_allstar\_v2.fits, while the distances and ages are from the GALAH\_DR3\_VAC\_ages\_v2.fits catalogue and were computed using the Bayesian Stellar Parameters estimator (BSTEP) method \citep{Sharma2018}. The dynamical parameters are from GALAH\_DR3\_VAC\_dynamics\_v2.fits, and were computed using the python package galpy \citep{Bovy2015} with BSTEP distances, astrometry from Gaia eDR3, and radial velocities from GALAH as inputs. All of these catalogues are downloadable from the GALAH survey website \footnote{\url{https://www.galah-survey.org/}}.

We limited the sample to stars flagged as having a reliable measurement for the stellar parameters, elemental abundances, Gaia parameters, and a signal-to-noise ratio of greater than 30 following the recommendations from \citet{Buder2021}. In addition to these quality cuts, we made a cut on the error in the distance determination to ensure reliable orbital parameters, as well as a metallicity cut on the sample at [Fe/H] < $-0.4$, in order to target stars in the metallicity range of $\ocen$. The cuts applied to the data are summarized with the following block of python code:


\begin{lstlisting}
galah_main[
        (galah_main.flag_sp == 0)           &
        (galah_main.flag_fe_h == 0)         &
        (galah_main.flag_x_fe == 0)         &
        (galah_main.fe_h <= -0.4)           &
        (galah_main.snr_c3_iraf > 30)       &
        (galah_main.e_distance_bstep < 1.0) &
        (galah_main.ruwe_dr2 < 1.4)        
        ]
\end{lstlisting}

Where galah\_main is a Pandas DataFrame \citep{Reback2020} containing the full Galah DR3 data set, and flag\_x\_fe are the respective flags for each element passed in as an input to the t-SNE analysis (where x = [$\alpha$, O, Na, Mg, Al, Si, K, Ca, Sc, Cr, Mn, Y, Ba]). This reduced the initial sample of 588,571 stars to 44,936. The GALAH DR3 data has a magnitude limit of $V \lesssim 14$, which means that red giant stars in the sample will probe out to a distance of no more than $\sim 8 \,\mathrm{ kpc}$. Indeed, an inspection of the distribution of distances (from the GALAH VAC, computed using BSTEP) for the GALAH sample shows two peaks at 1 and 2.5 kpc representing the disc and nearby halo and a tail extending out to 8 kpc (this distribution is shown in the grey histogram of the middle panel of Figure \ref{fig:hist_omega_cen}). Using parallax measurements from Gaia, \citet{Soltis2021} measure a geometric distance to $\ocen$ of $d_\odot = 5.24 \pm 0.11 \,\textrm{kpc}$ \citep[see also][ for distances derived with different methods, but yielding similar results]{Baumgardt_Hilker2018, Braga2018, Baumgardt2019}. The Fimbulthul stream spans a distance range of 2.4 - 7.2 kpc \citep{Malhan2022}, making this an appropriate sample for characterizing $\ocen$ and its associated structures.



\subsection{Pre-processing the data}
\label{sec:preprocess}


\begin{figure*}
	\includegraphics[width=\textwidth]{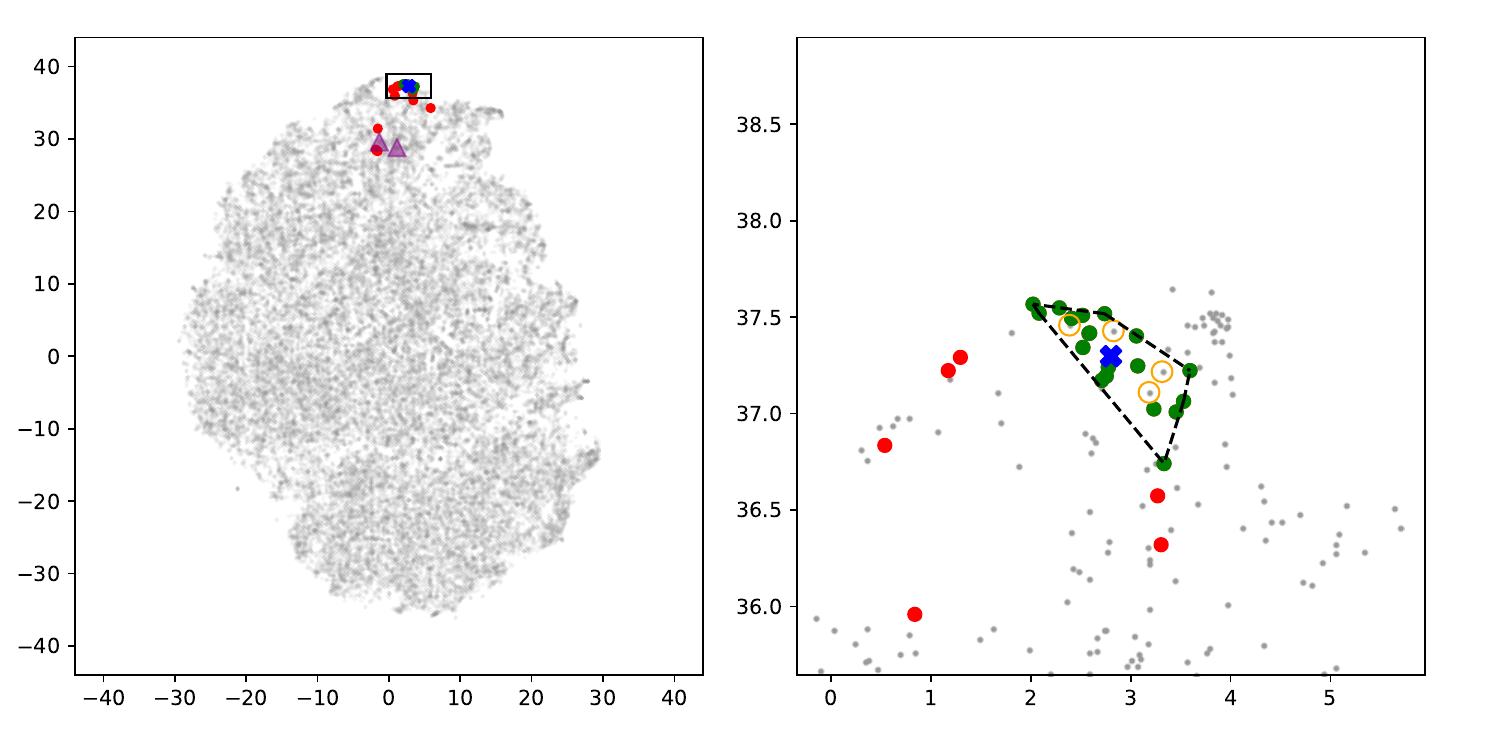}
    \caption{Left panel: t-SNE latent space projection for a single fiducial run of the $\ocen$ sample stars. Right panel: zoom in on the region containing the $\ocen$ candidate stars. Red and green points show literature members of $\ocen$, with green being the remaining stars after the sigma clipping procedure, and blue marking the mean coordinates of the centre point of this group. The orange circles show the selected candidates in the vicinity of the known literature members. Purple triangles are stars identified to be tidally ejected from $\ocen$ in \citealt{Simpson2020}. Note, the axes are dimensionless and do not represent any physical quantity, and distances between points should be interpreted as a relative representation of how clustered stars are in the high-dimensional space.}
    \label{fig:tsne_omega_cen_zoom}
\end{figure*}


One limitation to dimensionality reduction using t-SNE is that the input data cannot have any missing values. We therefore limited the number of elements included in the analysis, because although GALAH DR3 provides up to 30 elemental abundance measurements, abundances for some elements are not available for all stars, especially those that rely on few spectral lines and are difficult to measure. In addition, we only included stars for which the input abundances had a reliable measurement (flag\_x\_fe == 0) and omitted upper limits (flag\_x\_fe == 1). In order to select which abundances to include, we investigated the number of stars in the sample as a function of including different combinations of abundances (see figure 3 in \citet{Buder2022} for a visualization of this), while also taking into account which elements are important for identifying $\ocen$ stars.

In total, we included 13 elements and seven dynamical quantities for a total of 20 input parameters: $[\alpha/\mathrm{Fe}]$, [O/Fe], [Na/Fe], [Mg/Fe], [Al/Fe], [Si/Fe], [K/Fe], [Ca/Fe], [Sc/Fe], [Cr/Fe], [Mn/Fe], [Y/Fe], [Ba/Fe], $J_z$, $L_z$, $J_R$, energy (E), eccentricity, galactocentric pericentre distance (R$_{\mathrm{peri}}$), and galactocentric apocentre distance (R$_{\mathrm{apo}}$). We also conducted the analysis using different combinations and subsets of these parameters, including just kinematics and just chemistry, and found that including all of the above parameters was the most effective selection in terms of locating the $\ocen$ literature stars in a coherent region of the latent space.

In order to decide which chemical abundances to include, we experimented with many different combinations, and did our best to choose the set of elements which maximized the amount of chemical information input into the t-SNE, while minimizing the number of stars that were removed from the sample due to missing measurements. As a new approach in this work (compared to \citealt{Kos2018}), we also include orbital parameters as input into the t-SNE. In order for these to be useful for finding stars from the same cluster, we use the full 6D phase-space information, and input only quantities that are conserved along the orbit and over long periods of cosmic time. For example, the action-angle variables ($J_R$, $J_z$, and $J_{\phi}$) are extremely useful in this regard, as they are constants of motion when computed in an axisymmetric potential that is approximated as a Stäckel potential \citep{Binney2012}. Note, the azimuthal action ($J_{\phi}$) is equivalent to the z-component of the angular momentum, $L_z$, in an axisymmetric potential. These parameters therefore provide a very clear characterization of the orbits of stars in the Milky Way. Furthermore, the orbital energy, eccentricity, R$_{\mathrm{peri}}$, R$_{\mathrm{apo}}$ are characteristic parameters of the orbit and should also be conserved in an axisymmetric potential. We do not include other parameters such as tangential/radial velocity and proper motions as input, as these are subject to complex projection effects which can result in stars just a few degrees away from the cluster centre to have a vastly discrepant projected motion as compared to the cluster centre (see for example figure 2 from \citealt{Meingast2021} and figure 14 from \citealt{Bouma2021}). 

Before feeding these parameters into the t-SNE, we standardized each of them, meaning that each distribution was scaled to have a median of 0 and a standard deviation of 1. We did this by subtracting each value by the mean, and dividing by the standard deviation. This works well for parameters that are distributed symmetrically and whose distribution follows an approximate Gaussian, such as most of the chemical abundances, and some of the kinematic parameters. For $J_z$, $J_R$, and [Fe/H] we have asymmetric and monotonic distributions, and the standardization procedure still shifts them to have a median of 0 and a standard deviation of unity. This is not ideal, but it is important to treat these parameters in the same way so that each input parameter has an equal magnitude when interpreted by the t-SNE. 
In order to take into account the observational uncertainties and feature importance of the parameters, we then multiplied each abundance parameter by a weighting factor proportional to the measurement uncertainty. We adopted the weights used in \citet{Kos2018}, where scatter among stars in different clusters were measured for each element. Elements were then placed into four groups, based on how well measured a given element was in GALAH. For example, [Ba/Fe] and [K/Fe] had the most scatter, and were therefore given a weight of 0.25, [Mg/Fe] and [Ca/Fe] were slightly better and were given a weighting of 0.5, [Fe/H], [Ti/Fe], and [Cr/Fe] were the best measured and therefore given a weight of 2.0. All other elements were given a weight of 1.0. All of the weights and the details of their computation are provided in table 1 of \citet{Kos2018} and the associated text. We additionally have [Y/Fe] and [Mn/Fe] in our sample, to which we gave weights of 0.25 and 1.0, respectively. The kinematic parameters were all given a weight of 2.0, as this empirically improved the clustering of $\ocen$ stars in the latent space, indicating the importance of these kinematic parameters for selecting $\ocen$ stars from the background. 

The MultiCoreTSNE package requires two parameters to be initialized, n\_jobs, which simply defines how many CPU cores are dedicated for the computation, and perplexity, which is a measure of the size of clusters to be identified in the latent space. We chose n\_jobs = 4 and perplexity = 30. We experimented with different values of perplexity, and found that for an object like $\ocen$, a value of 30 produced the most reliable clustering of the literature members in the latent space. 

\subsection{Selection of candidates using t-SNE}

We ran t-SNE on the sample of 44,936 stars with the 20 chemical and kinematic parameters as inputs. The output is a 2D projection of the data, referred to as the latent space, which is shown in the left panel of Figure \ref{fig:tsne_omega_cen_zoom}. As with other manifold based clustering algorithms, t-SNE is initialized using stochastically assigned probability distributions, which results in a latent space projection that is slightly different each time when provided with the same input data. For this reason, we run the algorithm 100 times, and select candidates based on the stars which consistently fall in the region occupied by the known members of $\ocen$ from the literature.  

Figure \ref{fig:tsne_omega_cen_zoom} shows how the $\ocen$ candidate stars were selected from the latent space, for one t-SNE map out of the 100 iterations. In each mapping of the latent space, we used the positions of the confirmed $\ocen$ member stars from the literature to define the region in which we would expect to find $\ocen$ stars. Since most of these stars would clump in a given latent space mapping, we could identify stars that share similar chemical and dynamical properties with bonafide $\ocen$ members based on their proximity to this cluster. 

In order to reduce contamination, we performed a sigma clipping of the $\ocen$ literature stars (red points in Figure \ref{fig:tsne_omega_cen_zoom}) to remove outliers and select only the tightest cluster of the $\ocen$ stars. For this we used the astropy.stats sigma\_clip python package and found that clipping all stars two sigma away from the mean in both the x and y directions, and repeating this for three iterations resulted in the most consistent and pure selection of $\ocen$ literature stars in the latent space. The green points in the Figure show the stars that remained after the removal of outliers. We then drew a polygon with the outer edges defined by the shortest line encompassing all of the remaining green points using a SciPy implementation of a Convex-Hull, based on the Quickhull algorithm \citep{Barber1996}, and assigned a value of 1 for every star that fell in this region and 0 for every star that did not. Orange circles in Figure \ref{fig:tsne_omega_cen_zoom} show the selected candidates for this particular t-SNE iteration. We repeated this procedure 100 times, and then summed up the number of times a given star was labelled to be in the expected region of $\ocen$. We define this value as N$_{100}$, and will be using this notation for the remainder of the paper. 

\begin{figure}
	\includegraphics[width=\columnwidth]{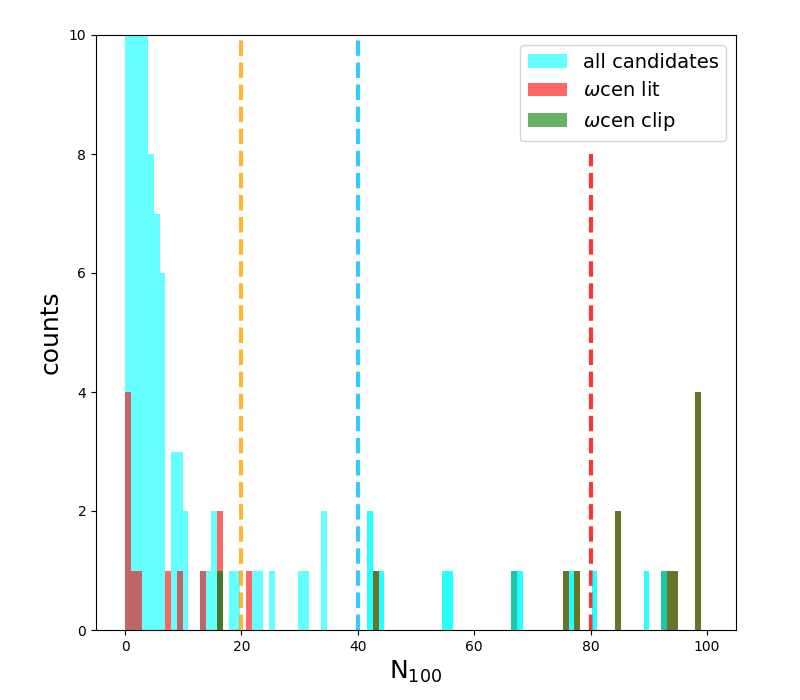}
    \caption{Distribution of N$_{100}$ values. The Cyan distribution show the entire sample, while red and green show the $\ocen$ literature stars, with the green stars corresponding to the selected green points in Figure \ref{fig:tsne_omega_cen_zoom}.}
    \label{fig:candidate_prob}
\end{figure}

It is important to note that N$_{100}$ does not directly translate to a probability of membership. For example, a candidate with N$_{100}$~=~20 does not mean that it has a 20\% chance of being a member of $\ocen$, but should rather be interpreted as a more likely than a candidate with N$_{100}$~=~10, but less likely than a candidate with N$_{100}$~=~30. The reason for this is because for each iteration of the t-SNE, we assign a point to a star that falls in the selected region in a binary fashion, such that there is no consideration that other stars may be close to having been counted, but fell just outside the region. Therefore, the the absolute value of N$_{100}$ is dependent on the choices made when defining the region around $\ocen$ literature stars. In order to check this, we ran the analysis performing only two iterations of the sigma clipping (which still makes a reasonable selection around the $\ocen$ literature stars) and found that the number of candidates with N$_{100} \geq$ 20 increased from 23 to 54. In fact, comparing the two samples, the N$_{100} \geq$ 20 group from the more strict selection was very similar to the N$_{100} \geq$ 40 group from the less strict selection, but the selection which contained more candidates also contained more true contaminants. We therefore decided to apply a relatively strict selection in the latent space and then include candidates with a lower N$_{100}$ threshold, as we believe this provides the cleanest selection of $\ocen$ candidate stars. This is further supported when looking at the distribution of N$_{100}$ values for the sample in Figure \ref{fig:candidate_prob}, showing a large number of contaminants that were selected a few times, with a sharp cut-off at N$_{100} \sim 10$. The dashed vertical lines in the figure show the selections at N$_{100}$ = 20, 40, and 80, which will make up the groups for the analysis in the rest of the paper.

\subsection{Generation of mock stellar streams}
\label{sec:mockstream_generation}

To predict the observable distribution of tidally disrupted debris from the interaction of $\ocen$ with the Milky Way, we simulated a suite of mock stellar streams using the Galactic and gravitational dynamics mock stream Python package \citep[Gala;][]{gala2017}. We used the MilkyWayPotential provided in the gala package, which is based on the four component MWPotential2014 described in \citet{Bovy2015}. This potential is made up from a combination of a Hernquist bulge and nucleus \citep{Hernquist1990}, a Miyamoto-Nagai disc \citep{Miyamoto_Nagai1975}, and a spherical NFW profile \citep{Navarro1997} to represent the dark matter halo. The parameters used for each component are summarized in Table \ref{table:stream_params}. Using this potential for the Milky Way, and a Plummer potential to account for the self-gravity of this in-falling cluster, streams were generated by computing the orbit of the cluster and allowing mass to be ejected through tidal interactions according to the prescription described in \citet{Fardal2015}. The streams are generated using backward orbital integration, first integrating the orbit backwards to the specified accretion time, and then allowing the cluster to evolve forward along the orbit with the constraint that the final position and velocity of the cluster match that of the present day location of $\ocen$. We allowed the initial stellar mass of $\ocen$ to vary from $1 \times 10^6\, \mathrm{M}_\odot$ to $1 \times 10^7\, \mathrm{M}_\odot$, and the simulation evolution time to vary from 1 to 12 Gyr to examine the changes in the expected morphology and evolution of the stream as a function of these two free parameters.


\begin{table}
\caption{Parameters for the Milky Way potential model used to generate mock stellar streams.}
\label{table:stream_params}
\begin{tabular}{lll}
Component & Parameter                & Value                                 \\
\hline
Bulge     & Bulge power-law exponent & -1.8                                  \\
          & Bulge cut-off radius     & 1.9 kpc                               \\
          & Mass                     & $5 \times 10^9\, \mathrm{M}_\odot$    \\
\hline
Nucleus   & Radius                   & 70 pc                                 \\
          & Mass                     & $1.7 \times 10^9\, \mathrm{M}_\odot$  \\
\hline
Disc      & Scale length             & 3 kpc                                 \\
          & Scale height             & 280 pc                                \\
          & Mass                     & $6.8 \times 10^{10}\, \mathrm{M}_\odot$ \\
\hline
Halo      & Scale radius             & 16 kpc                                \\
          & Mass                     & $5.4 \times 10^{11}\, \mathrm{M}_\odot$
\end{tabular}
\end{table}

In order to make a realistic comparison with the data, we also took into account the three dimensional footprint of the GALAH data, since it is not possible to detect candidates in regions where there are no observed stars in GALAH. To accomplish this, we applied a simplified selection function to mimic the three dimensional sky coverage of GALAH, by first binning the GALAH sample into equal area healpixels with NSIDE = 32. This corresponds to a physical separation of $\sim1.8 ^{\circ}$ between pixels, which is about the same size as a GALAH field ($\sim2^{\circ}$). We further binned each healpixel into distance bins of one kpc, and computed the ratio of GALAH stars in each distance bin to the total number of GALAH stars at all distances in the given healpix. We then down-sampled the number of stars for the mock stellar stream in each corresponding distance bin such that the relative ratio of stars per distance bin to total (for a given healpix) was the same for both GALAH and the mock stream. If there were too few stars in a given bin for the mock stream then all were taken, and in cases where the expected number of stars was between 0 and 1, we generated a random number between 0 and 1 and when this number was less than the number of expected stars we took one star from the mock stream, and when it was larger we took none. The result is a down-sampled mock stellar stream that matches the relative sky and distance distributions of the GALAH survey. Figure \ref{fig:l_b_omega_cen} shows the result of this procedure on the mock stream (coloured points), with the panels in the left column showing the raw output from the stream simulations and the panels in the right column showing the streams after down-sampling to match the GALAH footprint. The relative numbers of stars is dependent on the particular stream parameters used, since the star ejection rate is a constant parameter and therefore the final number of stars in the stream is proportional to the evolution time. To give an idea of how many stars are removed during the masking procedure, for the stream with $4 \times 10^6\, \mathrm{M}_\odot$ evolved for 8 Gyr (bottom row of Figure \ref{fig:l_b_omega_cen}) the number of stream particles is reduced from 16,002 (left panel) to 506 (right panel).

\section{Results}
\label{sec:results}

\begin{figure*}
	\includegraphics[width=\textwidth]{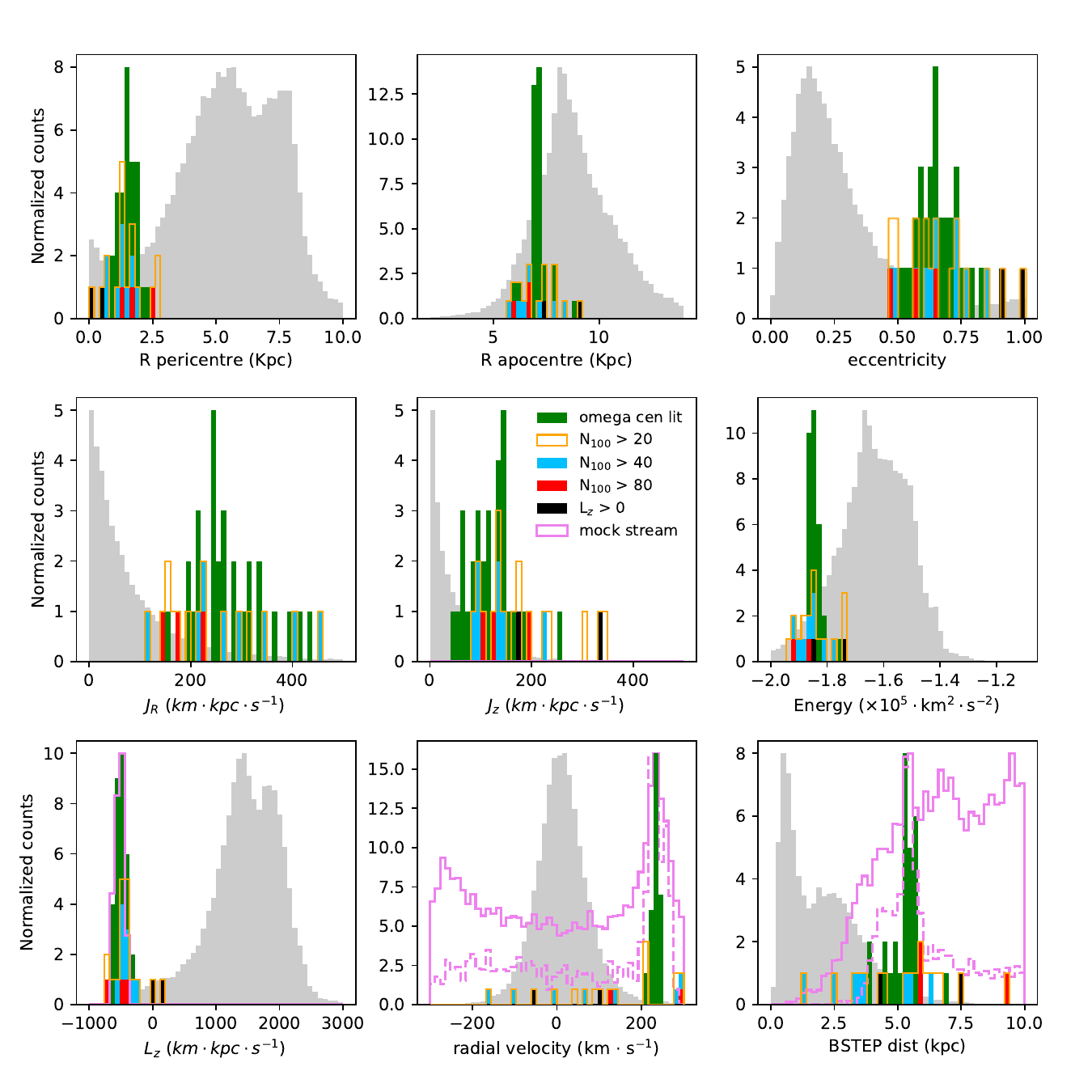}
    \caption{Distributions of kinematic parameters. The grey distribution is the GALAH sample, green is the literature members of $\ocen$, violet are the simulated mock stream stars with M = $4 \times 10^6\, \mathrm{M}_\odot$ and t$_\mathrm{acc}$ = 8 Gyr ago (solid) or t$_\mathrm{acc}$ = 1 Gyr ago (dashed), and $\ocen$ candidate stars selected from t-SNE with N$_{100} \geq 20$ (orange), N$_{100} \geq 40$ (blue) and N$_{100} \geq 80$ (red). Stars with near 0 or positive $L_z$ are flagged in black. Each parameter and how it was computed is described in the text in Section \ref{dynamical_analysis}.}
    \label{fig:hist_omega_cen}
\end{figure*}

\subsection{Validation of candidates}

In total, we found 18 candidates with N$_{100} \geq$ 20. Given that we identified these candidates using an unsupervised algorithm, it was necessary to retroactively validate that the selected stars have kinematic and chemical properties that are consistent with what is expected for stars stripped from $\ocen$. 

Orbital quantities computed using the full 6-d phase-space information, such as angular momentum and energy, should remain constant independent of the orbital phase. It should also be the case that these quantities remain coherent with the parent cluster over long periods of cosmic time, although it has been suggested that some properties of stream stars, such as their energy, can change over time due to tidal interactions which depend on the mass ratios of the in-falling satellite and host Galaxy \citep{Ibata2019a}. Other kinematic parameters such as radial velocity and proper motions can vary substantially as a function of orbital phase due to projection effects across the sky \citep{Meingast2021,Bouma2021}. It is for this reason that we do not include these parameters as input into the t-SNE analysis.

\subsubsection{Dynamical analysis}
\label{dynamical_analysis}

Figure \ref{fig:hist_omega_cen} shows the distributions of the selected candidates and literature $\ocen$ stars for various kinematic parameters. The quantities shown in each of the panels are as follows: Galactocentric pericentre radius, Galactocentric apocentre radius, eccentricity, radial action ($J_R$), vertical action ($J_z$), energy, z-component of the angular momentum ($L_z$) radial velocity (from GALAH), and the heliocentric distance (from GALAH VAC computed using BSTEP).

The grey histogram shows the distributions of all stars in GALAH with [Fe/H] < $-0.4$, and which passed the quality cuts described in Section \ref{sample_selection}. The violet histogram represents the distribution of stars from one of the mock stellar streams (M = $4 \times 10^6 \mathrm{M}_\odot$ and t$_\mathrm{acc}$ = 8 Gyr ago), and is plotted for each parameter which is available. Here we differentiate the candidates into three different groups based on their N$_{100}$ score, to separate candidates based on the likelihood that they can be associated to $\ocen$. The $\ocen$ literature members are plotted in green, and the candidates from the t-SNE analysis are plotted in red for N$_{100} \geq 80$, blue for N$_{100} \geq 40$ and orange for N$_{100} \geq 20$.

Looking at the bottom left panel of the $L_z$ distribution, $\ocen$ is clearly distinguishable from the bulk Milky Way population. The majority of stars in the sample are disc stars, orbiting in a prograde direction ($L_z$ > 0) around the Galactic center. There is a small tail in the grey histogram at negative $L_z$, representing the stars with a retrograde orbit. These stars have all likely been accreted, as an opposite rotational velocity to the disc is not likely for \textit{in situ} stars. All of the $\ocen$ literature stars have a large and negative $L_z$, and all of the candidates with N$_{100} \geq 20$, as well as all of the mock stream stars are all clumped in this region of the plot. There are only two candidate stars with positive or near zero values for $L_z$, which we flag in all panels of the figure in black. The panel showing the energy distribution is quite similar, as almost all $\ocen$ literature stars and candidates clustering tightly at high energy, and with one of the lowest energy stars flagged with $L_z \sim 0$ .

The two left panels in the middle row show the other two components of the action vector in cylindrical coordinates, $J_R$ and $J_z$. In both of these distributions we also see that $\ocen$ stars are located in the tail of the distribution, at large values of $J_R$ and $J_z$, respectively. However, the distributions of literature stars and candidates are broader, and overlap more with the Milky Way stars than they do for the $L_z$ distribution.

From the top row of Figure \ref{fig:hist_omega_cen}, we can see that there is also some selection power in the pericentre, apocentre and eccentricity parameters. The $\ocen$ literature stars seem to be characterized by a large difference in their pericentre and apocentre distances, which is also evident in the concentration of $\ocen$ literature and candidate stars at high eccentricities.

Interestingly, the two stars with $L_z \sim 0$ also pass very close to the Galactic centre at pericentre, and have highly eccentric orbits. Taking into account the high $J_z$ for these stars, these may well be accreted stars on a somewhat peculiar, near vertical orbit passing through the Galactic center. Another possibility is that these highly unusual orbits are a result of artefacts from the numerical integration, due to inaccurate orbital properties for these stars. In any case, they are most likely not $\ocen$ stars that have been ejected from the cluster, at least not from normal tidal interactions with the Milky Way. Therefore, we remove these two stars from the analysis when we compare the on sky distribution of stream candidates to mock stream stars later in the paper in Section \ref{subsec:constraining_mass_and_acc_time}.

The bottom right panel shows the distribution of heliocentric distances. As expected, the $\ocen$ literature stars form a group around 5.2 kpc, but with a slight spread due to uncertainties in the individual distance measurements. The candidate stars are also grouped in the range of three to seven kpc, with a few outliers outside of this range. The distribution for the mock stream stars peaks at the distance of $\ocen$, and then remains flat out to larger distances. For the stream simulations, the number of stars at larger distances increases with longer evolution times. This is depicted in the figure by the dotted violet line, which shows a stream with an evolution time of 1 Gyr, as compared to the solid line which shows the same for an evolution time of 8 Gyr.



Finally, the bottom panel of the middle row shows the distribution in radial velocity. The candidates are distributed rather uniformly across the range of parameter space, with a slight peak around the $\ocen$ literature stars, as is predicted from the distribution of mock stream stars. Again, the mock stream distributions show that the longer the evolution time, the more stars that are expected to have radial velocities that are significantly different than the main cluster. This can be attributed to projection effects as ejected stars distribute more and more across the sky (and along their orbits), they will have more or less of their velocity in the radial direction (and a change in the magnitude of their velocities proportional to their proximity to peri- or apo-centre).

In addition to the histograms shown in Figure \ref{fig:hist_omega_cen} we have also provided some additional plots of kinematic spaces in the Appendix. The E-$L_z$ diagram is a commonly plotted parameter space in the literature when discussing accreted substructure. Figure 
\ref{fig:E_Lz_diagram} confirms the histograms in Figure \ref{fig:hist_omega_cen}, showing a tight concentration of energies and angular momenta for the $\ocen$ stars, with the t-SNE candidates occupying a slightly broader region of the parameter space and two outliers at near zero and positive $L_z$. Figure \ref{fig:rperi_rapo_plot} shows the same features as well, but with the outliers at low R$_{\mathrm{peri}}$. Figures \ref{fig:Lz_Jz_diagram} and \ref{fig:Jz_Jr_diagram} show different configurations of the action space, which demonstrates that $\ocen$ stars do not stand out individually in $J_R$ or $J_z$, but in the combined $J_R$ vs $J_z$ space, they are clearly differentiable from the bulk population. The pmra vs pmdec plot (Figure \ref{fig:pm_diagram}) shows the regions occupied by $\ocen$ literature and Fimbulthul stars, as well as a region defined in \citet{Soltis2021} to contain $\ocen$ cluster stars. The t-SNE selected candidates do not obviously cluster near them, but this is not surprising given the distance between them on sky, which causes the proper motions to be strongly affected by projection effects.



\begin{figure*}
	\includegraphics[width=\textwidth]{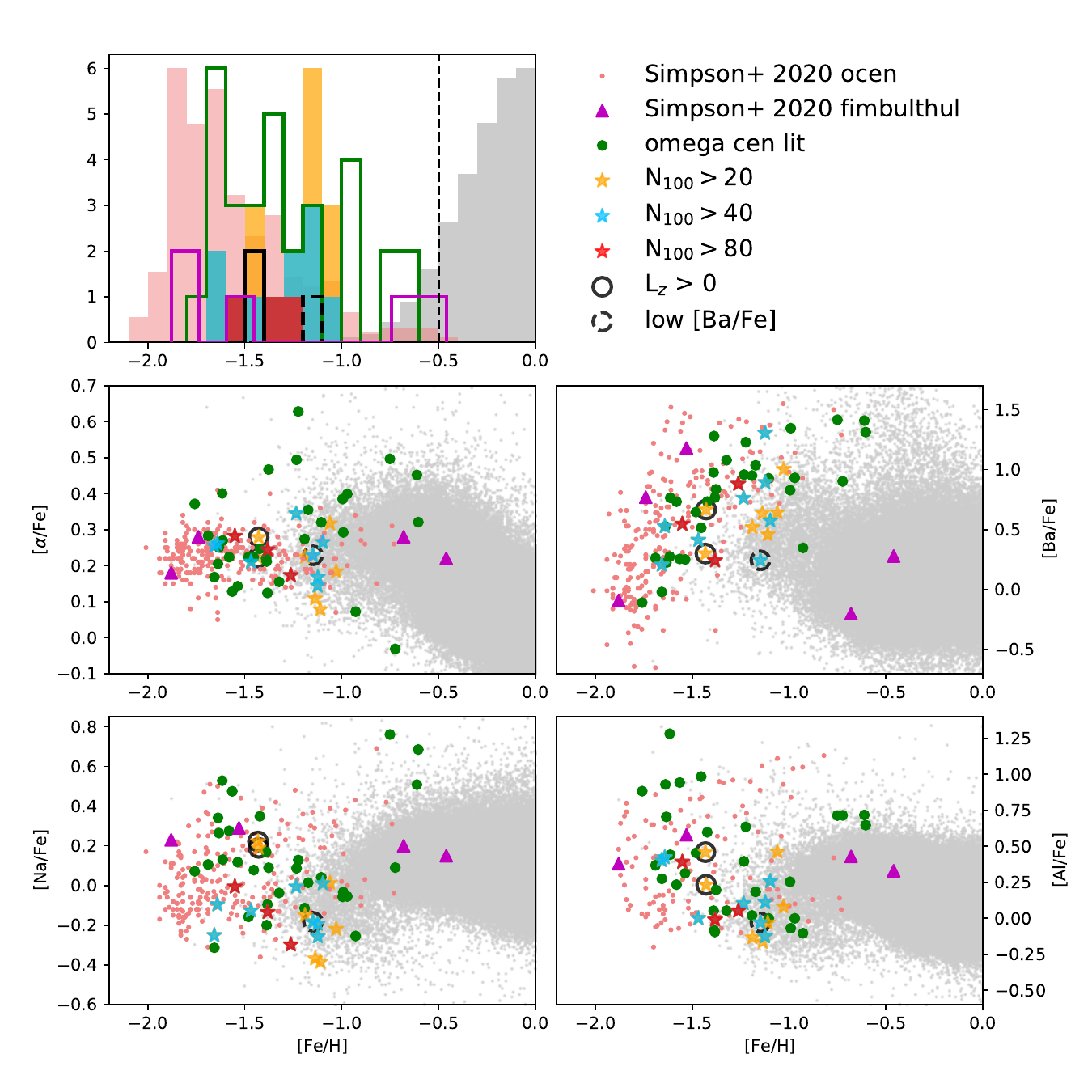}
    \caption{Chemistry of $\ocen$ candidates and confirmed cluster members. Top left: histogram of [Fe/H] distributions. Other panels are scatter plots for various elemental abundances, [X/Fe] vs [Fe/H]. Orange, blue and red star symbols are candidates with N$_{100} \geq$ 20, 40 and 80, respectively. Grey points are a sub-sample of the GALAH data applied and represent the bulk MW halo and disc populations. Pink points and purple triangles are the $\ocen$ literature sample and chemically selected Fimbulthul $\ocen$ candidates from \citet{Simpson2020}, respectively. Green points are $\ocen$ literature members from the GALAH sample in this work. Stars with near 0 or positive $L_z$ are circled in black.}
    \label{fig:ocen_chemistry}
\end{figure*}

\subsubsection{Chemical analysis}

Figure \ref{fig:ocen_chemistry} shows the distributions of the selected candidates and literature $\ocen$ stars in various chemical abundance planes. As can be seen in the top left panel, all of the candidates and literature members of $\ocen$ have low values of [Fe/H]. For the candidates, this is partially by design, since the sample was cut at [Fe/H] < $-0.4$. However, almost all of the candidates occupy the range of $-2 < \mathrm{[Fe/H]} < -1$, which is contained within the metallicity range of $\ocen$ reported in the literature \citep[e.g.][]{Johnson2010}. 


The first panel in the second row shows [$\alpha$/Fe] vs [Fe/H], where [$\alpha$/Fe] is a weighted average of independent measurements of alpha elements provided in the GALAH DR3 catalogue. 
There is quite a bit of scatter here, especially for the $\ocen$ literature stars (green points). The t-SNE candidates all seem to be alpha enhanced, but only slightly, such that they would not be differentiable from halo stars of the same metallicity.


The third panel show the s-process element Barium. For $\ocen$ stars, we expect [Ba/Fe] to show a significant enhancement with respect to Milky Way field stars at [Fe/H] $\sim -1$, and then decreases down to solar levels at lower metallicities of [Fe/H] $\sim -2$ \citep{Norris_DaCosta1995b, Smith2000, Majewski2012}. This is clearly seen for the literature stars (pink and green points), but also holds for most of the candidate stars, especially those with N$_{100} > 40$ (blue and red stars). 
There is a small group of candidates at [Fe/H]~$\sim -1.1$ which seem to be lower in [Ba/Fe] with respect to this trend. However, this not a very significant depletion (with the exception of the one blue star at [Ba/Fe] $\sim$ 0.2), and may simply be attributable to a spread in [Ba/Fe] at this metallicity (there are also salmon and green $\ocen$ literature points near this region) or the measurement uncertainties from the GALAH abundances, which are known to be large for [Ba/Fe] \citep{Kos2018}. We flag the star with the lowest [Ba/Fe] (shown by the dashed black circle in Figure \ref{fig:ocen_chemistry} and remove it from the candidate sample for the analysis in the rest of the paper.

As for [Na/Fe] and [Al/Fe], the candidate stars appear to have systematically lower abundances of [Na/Fe] than the literature members of $\ocen$, as well as the candidates from \citet{Simpson2020} (purple triangles). They are, however, not more depleted than some $\ocen$ stars with lower abundances in these elements, and are within the expected spread of [Na/Fe] abundance for $\ocen$ stars.

The two candidate stars flagged as outliers in $L_z$ space (black circles) seem to be chemically very compatible with the $\ocen$ literature stars. This is true for nearly all of the abundance spaces. This is not surprising, given that they were identified by the t-SNE as $\ocen$ stars, and therefore must have been convincingly similar in their abundance parameters to be clumped in the latent space, despite having discrepant kinematics.

The distributions for the remaining elements used as inputs into the t-SNE are also provided in Figure \ref{fig:appendix_chemistry} in the Appendix. They also show a good general agreement between the $\ocen$ literature stars and the t-SNE selected candidates. We also investigated various anticorrelations suggested in the literature to exist in $\ocen$ stars, including Na--O, Al--O, and Mg--K, as well as possible [Ba/Eu] enhancement. We did not find any noticeable anticorrelations in light element abundances, nor did we find any particular evidence of s-process enhancement for either the $\ocen$ literature samples or the t-SNE selected candidates, suggesting that these data are not sufficient to show these trends. These plots are included in the Appendix for the curious reader, Figure \ref{fig:anti_corr_diagram}.




\begin{figure*}
	\includegraphics[width=\textwidth]{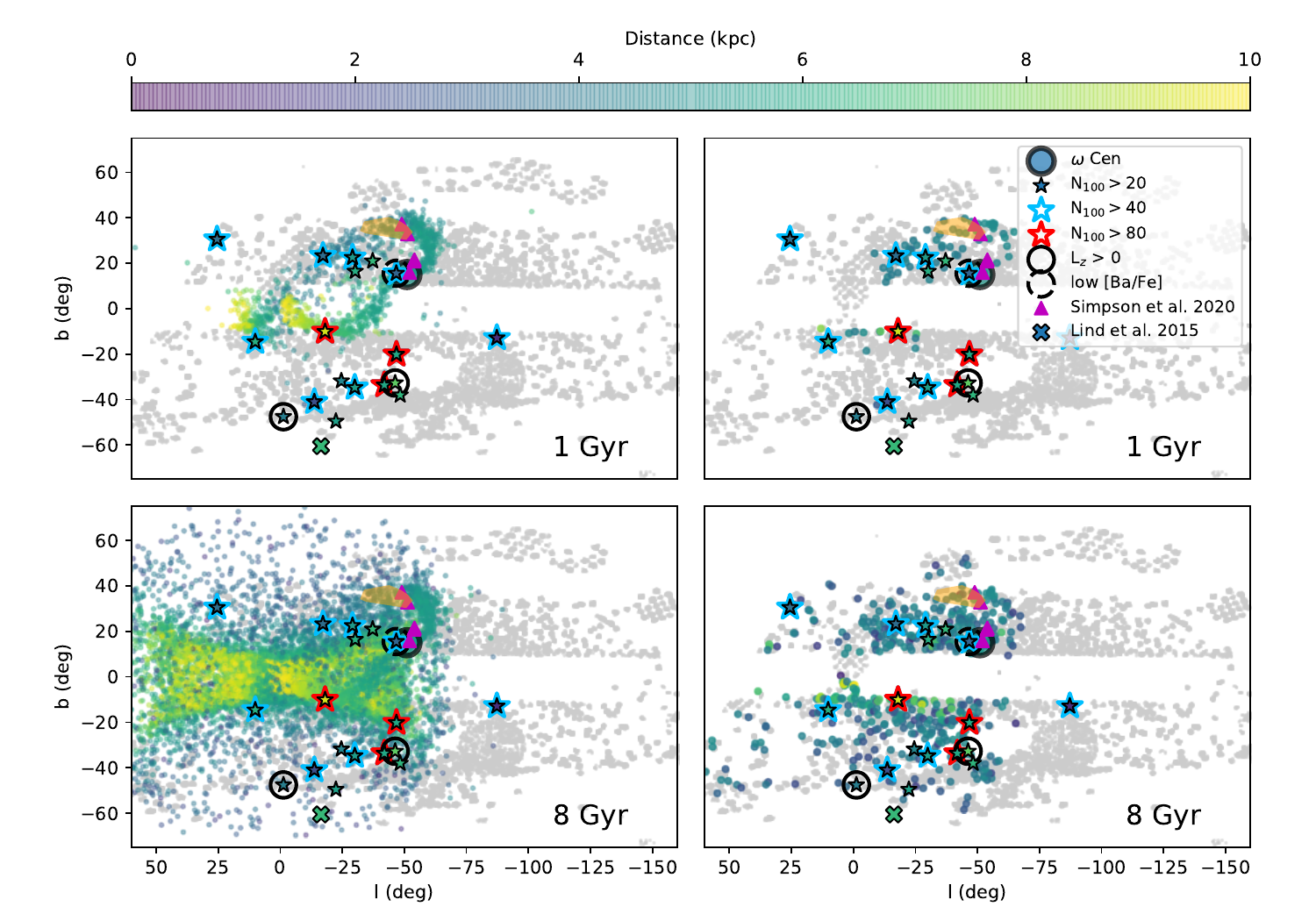}
    \caption{Spatial distribution on the sky of candidates for mock streams with a progenitor mass of M = $4 \times 10^6\, \mathrm{M}_\odot$. The top row is for mock stream simulations that evolved for just 1 Gyr, and the bottom row was allowed to evolve for 8 Gyrs. The left panels show all stars from the mock stream simulations with d < 10 kpc, and the right panels show the mock stream stars after applying the GALAH observational imprint. Grey points are all of the GALAH stars with [Fe/H] < $-0.4$, black points are Fimbulthul members from \citet{Ibata2019a}, the green X is the star from \citet{Lind2015}, purple triangles are $\ocen$ candidates from \citet{Simpson2020} and the coloured circle marks the location of $\ocen$. Stars from the mock stream are shown as dots and star symbols show the candidates selected from t-SNE with N$_{100} \geq$ 20, with stars flagged in $L_z$ space circled in black. All stars and points are coloured according to their heliocentric distances, shown by the colour bar.}
    \label{fig:l_b_omega_cen}
\end{figure*}

\section{Discussion}
\label{sec:discussion}

\subsection{Constraining the initial mass and accretion time of $\ocen$}
\label{subsec:constraining_mass_and_acc_time}

In this section, we use mock stellar stream simulations to determine which stream properties best reproduce the observed distribution of the $\ocen$ candidate stars. To accomplish this, we compute mock stellar streams using the Gala mockstream python package \citep{gala2017} over a grid of different masses and accretion times, covering the ranges $10^6 \,\mathrm{M}_\odot \leq \mathrm{M} \leq 10^7 \mathrm{M}_\odot$ and 1 Gyr $ \leq \mathrm{t}_\mathrm{acc} \leq$ 12 Gyr. In total, this constitutes 48 mock stream simulations.

 Two of these mock streams are shown in Figure \ref{fig:l_b_omega_cen} (coloured points in the left panels), along with the spatial distribution on the sky of the $\ocen$ candidates, depicted as star symbols and coloured by their heliocentric distances. The panels on the right show the streams after the application of the GALAH survey observation pattern, as described in Section \ref{sec:mockstream_generation}. The most striking feature of the $\ocen$ candidate stars is that they are located at large distances from the main body of $\ocen$. These stars are also predominantly located at $|l| < 65$, i.e. in the half of the sky towards the galactic centre with respect to the main body of $\ocen$. The top left panel of Figure \ref{fig:l_b_omega_cen} shows a stream simulation with M = $4 \times 10^6 \mathrm{M}_\odot$ but that was only allowed to evolve for 1 Gyr. The stream is still very coherent, and very few stars have been ejected towards the solar neighbourhood such that they can be observed by GALAH (right panel). For the same mass of stream, but allowing it to evolve for 8 Gyr (bottom left panel), we see that much more mass from the stream has been dispersed and there are many more stars observable by GALAH, making a much better match to the observations.
 


In a recent study, \citet{Ibata2019a} used $N$-body simulations of a stellar stream to show a connection between $\ocen$ and the Fimbulthul stream, a newly discovered stream using the STREAMFINDER algorithm \citep{Ibata2019b}. The region containing the candidate members of this stream from \citet{Ibata2019a} is shown in orange, just above the location of the main body of $\ocen$ in Figure \ref{fig:l_b_omega_cen}. In another study, \citet{Simpson2020} identified a sample of stars in Fimbulthul that they chemically linked to $\ocen$, using abundances from the GALAH survey. These stars are shown as the purple triangles in the figure. Finally, \citet{Lind2015} found a star in the Gaia ESO Survey which showed a highly peculiar chemical abundance pattern consistent with a globular cluster origin, and with a kinematic analysis they determined that $\ocen$ was the most likely cluster for it to have been ejected from. This star is demarcated with a green X in Figure \ref{fig:l_b_omega_cen}, and its large distance from the main body of $\ocen$ is indicative of how far stripped stars can be found from their parent cluster. However, it should be noted that the proper motions used to calculate the orbit of this star were from the Tycho-Gaia astrometric solution (TGAS) from Gaia DR1, and the proper motion measurements have changed significantly since then in Gaia DR3, so it is unclear if this star would still be associated with $\ocen$ if the orbit calculations were to be recomputed.

\begin{figure*}
	\includegraphics[width=\textwidth]{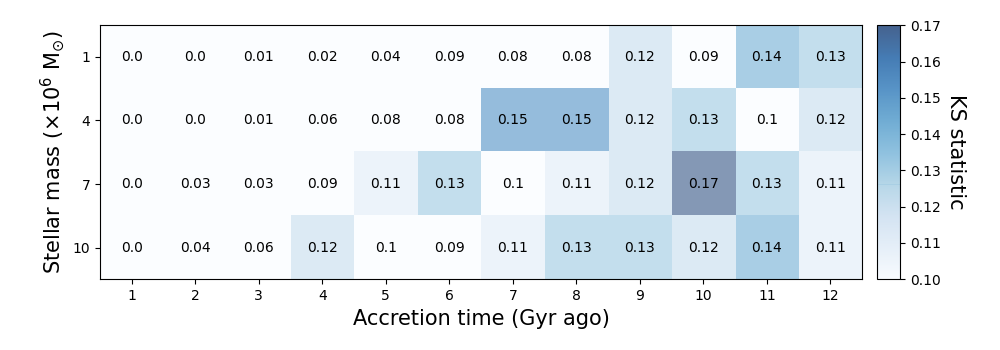}
    \caption{Grid of computed $p$-values from the 2D KS test for each combination of mass and evolution time of the mock stellar streams. Squares are coloured based on the $p$-value magnitude, and each square corresponds to the sky projections of $\ocen$ candidates with N$_{100} \geq$ 20 and mock stellar stream stars included in Appendix \ref{sec:appendix_A}.}
    \label{fig:KS_grid}
\end{figure*}

Although it is clear that the lower panels of Figure \ref{fig:l_b_omega_cen} are qualitatively a better match to the observed distribution of tidally stripped $\ocen$ candidates, we would like to explore a larger set of models and use a more quantitative measure to see if we can further constrain the initial mass and accretion time of $\ocen$. To this end, we took all 48 of the stream models after applying the observational footprint of GALAH, and performed a 2D Kolmogorov--Smirnov (KS) test\footnote{The code used for the 2D KS test was modified from \url{https://github.com/syrte/ndtest}}. A KS test statistically determines the probability that two distributions are consistent with being drawn from the same parent distribution. In other words, the test computes a KS statistic, or a $p$-value, which can be used to determine if the two samples have statistically different distributions. Larger $p$-values (p > 0.2) indicate that the samples could be drawn from the same distribution, while small $p$-values suggest that they are not. We computed KS statistics for each of the model mock streams and the candidate sample with N$_{100} \geq 20$. For this analysis, we excluded the stars with positive or close to zero L$_{z}$ and low [Ba/Fe], as well as the star at l = $-85$, reducing the candidate sample from 18 to 14 stars. The results are presented in Figure \ref{fig:KS_grid}. The corresponding on sky distributions in Galactic coordinates for each square in this grid are included as Figures \ref{fig:mock_stream_grid_1_6} and \ref{fig:mock_stream_grid_7_12} in Appendix \ref{sec:appendix_A} for reference.

From Figure \ref{fig:KS_grid}, we see that given the distribution of observed candidates with N$_{100} \geq 20$, none of the models reach the p = 0.2 threshold to be statistically significant, but nevertheless we can still use the $p$-values to rule out some models and suggest which which models may be favoured. For example, stream models which are only allowed to evolve for 3 Gyr or less, regardless of initial mass, have very low $p$-values which indicates the distribution of mock stream stars and observed $\ocen$ candidates are not compatible with being drawn from the same distribution. This can be interpreted as these mock stream models not having enough time to eject sufficient mass to account for the observed distribution of tidal debris. If the initial mass of $\ocen$ was similar to its current day stellar mass of M = $4 \times 10^6\, \mathrm{M}_\odot$ \citep{DSouza_Rix2013}, i.e. the fractional mass loss is very low, then this would place a constraint on the accretion time of $\ocen$ at t$_{\mathrm{acc}}$ > 7 Gyr. At slightly longer evolution times of 4 Gyr, the models with the highest initial mass of $\mathrm{M} = 10^7 \mathrm{M}_\odot$ show slightly higher $p$-values, and although these are not conclusive it still suggests that these models can not be ruled out. As the initial mass of the mock stream models increase, the plausible evolution times also increase, which can be seen by the step-like increase of the blue squares in Figure \ref{fig:KS_grid}. At all masses, the longer evolution times (t$_{\mathrm{acc}} > 9$ Gyr) also result in an inconclusive KS-test and also cannot be ruled out. Therefore, based on these data, we can only put a lower limit on the accretion time of $\ocen$, which is dependent on the initial mass.

Finally, if we consider that major mergers capable of heating up the thick disc and bringing in large satellites are likely to have happened early on in the formation of the Milky Way \citep{Helmi2018,Belokurov2018, Xiang_Rix2022}, and if $\ocen$ were to have been brought in along with such a merger then this would suggest that more ancient accretion times and a smaller progenitor mass would be favoured (i.e. t$_\mathrm{acc}\, \geq$ 7 Gyr). However, this scenario is still speculative, and further evidence is needed in order for it to be confirmed.

We also computed KS-test statistics for the samples with N$_{100} \geq$ 40 and 80 and included these in the Appendix. With N$_{100} \geq$ 40 the sample is reduced to nine stars, and shows approximately the same results as with N$_{100} \geq$ 20, but with higher $p$-values for nearly all models, and also favours the most massive progenitors, and longest evolution times, likely because the stars from these models are more uniformly distributed. For the N$_{100} \geq$ 80 sample, only three stars remain and due to the small sample size the KS-test is uninformative at all masses and accretion times.

\subsection{$\ocen$ as a stripped dwarf galaxy}

If we consider the scenario that $\ocen$ is the remaining core of a stripped dwarf Galaxy, then we would expect $\ocen$ to have been initially much more massive than the present day. \citet{Tsuchiya2003} explore the plausibility of a capture scenario for $\ocen$ by performing an $N$-body simulation of a dwarf Galaxy accretion onto a Milky Way like potential. They find that the progenitor which best reproduces $\ocen$'s present day position, mass, and orbital properties is a nucleated dwarf Galaxy with a Hernquist density profile and a total mass of 8 x 10$^9 \textrm{M}_{\odot}$. Furthermore, they show that a non-nucleated dwarf loses mass too quickly, and a density profile that is too cuspy does not lose mass quickly enough. They find that after only 3 Gyr, the orbit of their progenitor decays to a similar orbit as $\ocen$, and the mass loss is sufficient to reproduce a cluster with roughly twice the mass of $\ocen$.

In another study by \citet{Bekki2003}, they use a dynamical model to attempt to reproduce the orbital parameters and mass of $\ocen$ while also providing a plausible scenario for its chemical enrichment. They assume a mass fraction for the nucleus of $f_n = 0.05$ and a mass loss through stripping from tidal interactions of $f_\mathrm{lost} = 0.2$, such that they compute an initial mass for the cluster of 1.25 x 10$^8 \textrm{M}_{\odot}$. They then merge this $\ocen$ like progenitor with a Milky Way like Galaxy with a young Galactic disc (10 Gyr ago), and find that the stellar nucleus survives while the outer envelope is completely stripped. They also show that the nucleus does not survive and the cluster is completely disrupted after just 2.5 Gyr when they include a Milky Way bulge component, or a larger disc mass. Similarly, if the $\ocen$ progenitor has a smaller mass and density, it also does not survive to the present day. Both \citet{Bekki2003} and \citet{Tsuchiya2003} predict a significant amount of stripped material from $\ocen$, which consists of a retrograde and fairly metal-poor population, although they both state that the density of these stars are expected to be quite low in the solar neighbourhood \citep[see for example the right panel of figure 5 in][]{Bekki2003}.

In our stream models we do not consider the disruption of the main body of $\ocen$ itself, but rather just the distribution of ejected stream stars. Furthermore, we do not explore mass ranges as large as those used in the dynamical models of \citet{Bekki2003}, as more massive models did not converge for long evolution times when run in Gala. Nevertheless, taking the results of \citet{Bekki2003} into consideration, it seems that a scenario in which $\ocen$ was initially much more massive (with a dense central nucleus) and an early accretion time of t$_{\mathrm{acc}} > 10$ Gyr ago, is the most likely scenario to account for its present day orbital characteristics and chemical abundance signatures. However, based on the findings of \citet{Tsuchiya2003}, it is unclear if the progenitor of even a nucleated dwarf could survive for such a long time if the Galaxy had a more massive disc and bulge component. From the lower row of Figure \ref{fig:KS_grid}, the current findings place a constraint of t$_{\mathrm{acc}} > 4$ Gyr ago for a massive progenitor and do not allow for a distinction between earlier accretion times. Therefore, the specific accretion time of $\ocen$ remains an open question, but the scenario that it is an accreted core of a dwarf Galaxy is still highly alluring.

\subsection{Caveats of a static Milky Way potential}
As previously mentioned in Section \ref{sec:mockstream_generation}, the orbital parameters used as input into the t-SNE and the mock stellar stream simulations were computed using the \textit{MWPotential14} \citep{Bovy2015}, which is an axisymmetric, static Milky Way potential. Although this potential nicely models the current Milky Way mass distributions and includes a nucleus, bulge, disc and halo, it does not account for time-dependent, non-axisymmetric perturbations caused by a rotating bar or an infalling Large Magellanic Cloud (LMC). 

It has been suggested that a massive LMC (i.e. 1--2.5 $\times 10^{11}\,\mathrm{M}_{\odot}$ is capable of significantly perturbing the gravitational potential of the Milky Way \citep{Garavito-Camargo2019, Garavito-Camargo2021, Cunningham2020}, and $N$-body simulations taking into account an in-falling LMC have shown that it can significantly alter the orbits of other satellite dwarf galaxies \citep{Battaglia2022}. However, it should also be noted that some of the systems investigated in \citet{Battaglia2022} were barely affected by the presence of an LMC, and this was dependent on the proximity to the LMC and the MW reflex motion caused by the LMC's gravitational wake in the halo. Furthermore, these simulations assumed a massive LMC, and no consensus on the mass of the LMC has yet been reached in the literature.

Nevertheless, an evolving potential may also cause conserved orbital quantities such as $L_z$ and E to dissipate over time, resulting in a less coherent clustering in phase space for stars from a common progenitor. That being said, in this work, candidate stars were identified empirically by their clustering in latent space with known, bonafide members of $\ocen$, and these members were very clearly clustered in their kinematic and orbital parameters, suggesting that this particular cluster has not been strongly affected by non-axisymmetric perturbations. Furthermore, the conclusions in this work drawn from the comparisons to the mock streams are mild, and reflect the uncertainty present in the simplified potential used. Therefore, although it is outside of the scope of this paper, future work should implement comprehensive $N$-body simulations, which include a time dependent potential and an in-falling LMC in order to properly model the evolution of both the tidally ejected stream stars and the the mock stream models.

\subsection{Connection to the Fimbulthul stream}

Using an $N$-body simulation of a model stellar stream, \citet{Ibata2019a} proposed $\ocen$ to be the parent cluster of the Fimbulthul stream. In the current work, all stream simulations -- independent of initial mass and accretion time -- showed substantial overlap with the region containing the Fimbulthul stars, thereby suggesting consistency with the results of \citet{Ibata2019a}. However, looking at the shaded orange region in Figure \ref{fig:l_b_omega_cen}, we do not find any candidate members directly contained within the Fimbulthul stream. Upon a close inspection of the bottom right panel in Figure~\ref{fig:l_b_omega_cen} (showing the mock stream after application of the observational footprint of GALAH) we see very few stream stars in the Fimbulthul stream region, demonstrating that this could be the result of an observational bias due to a lack of coverage in this part of the sky by GALAH, in combination with our selection criteria for the t-SNE input parameters. Indeed, the area just below the Fimbulthul region (at l = $-30$ and b = 25) shows a significantly higher density of mock stream stars than in the Fimbulthul region, and there are four candidates in this region.

In order to assess whether we would find stars in the Fimbulthul region with a different sample selection, we ran the t-SNE analysis using different combinations of input parameters. Using just kinematics and a few elemental abundances ([Fe/H], [$\alpha$/Fe], [Ba/Fe], and [Y/Fe]), found several stars in the region connecting $\ocen$ and Fimbulthul, as well as a few stars in the Fimbulthul stream, including the three stars identified in \citet{Simpson2020} as being ejected from $\ocen$. However, reducing the number of input parameters also increased the total number of candidates, and also the number of contaminants. This is why we ultimately decided to use the full complement of abundances described in Section \ref{sec:preprocess} for the main analysis of the paper.

\subsection{Contamination in the candidate sample}

We also consider the possibility that we could have chosen candidates nearby $\ocen$ literature stars in the t-SNE latent space that have serendipitously fallen in the selection region. Although this likely did occur for any given run of the t-SNE algorithm, we would expect different stars to scatter randomly into the selected region of the latent space for each iteration of the 100 runs, and thus most spurious contaminants to have low values of N$_{100}$. In order to check that this was indeed the case we also performed an additional test. Before running t-SNE, we first randomly shuffled the values of all of the input parameters such that each star had a random combination of parameters from other stars in the sample. We then appended on the $\ocen$ literature sample with the correct parameters, in order to identify the region in latent space where $\ocen$ stars should be located. After running the 100 iterations of the t-SNE on this sample we found no candidates with any N$_{100}$ values greater than zero. In other words, the $\ocen$ stars were highly efficiently selected in the latent space, and no random stars were mistaken to be candidates. The latent space projection associated with this test is shown in Figure \ref{fig:tsne_omega_cen_zoom_shuffle} in the Appendix.
 
To ensure that there were no systematic issues or unreliable abundance measurements, we inspected the GALAH spectra and synthetic spectral fits for the stellar parameter and [Fe/H] runs by eye for each of the t-SNE selected $\ocen$ candidates. No candidates were removed based on this check.


\subsection{Future prospects}
Given that this method relies upon high dimensional data and is computationally fast, it is very well-suited to be applied to upcoming large spectroscopic data-sets such as WEAVE \citep[]{Dalton18} and 4MOST \citep{deJong19}. Each of these surveys will observe millions of stars at high resolution allowing for the determination of many elemental abundances. When combined with the astrometric data from Gaia, this will be a very powerful multidimensional data set with which to find tidally stripped members of clusters across the entire sky.



\section{Conclusions}
\label{sec:conclusions}

In this work, we used t-SNE to perform dimensionality reduction on GALAH DR3 plus Gaia eDR3 data to find stars that have been tidally stripped from the star cluster Omega Centauri. After identifying these candidates, we validated them by looking in more detail at their individual kinematic and chemical abundance signatures, which removed two candidates with L$_z \sim$ 0, one candidate with low [Ba/Fe], and one candidate that was an outlier in Galactic longitude at l < $-75$, resulting in a final sample of 14 candidates with N$_{100} \geq$ 20.

We then ran a grid of mock stellar streams with a range of initial stellar masses from $10^6 \,\mathrm{M}_\odot \leq \mathrm{M} \leq 10^7 \mathrm{M}_\odot$ and evolution times of $1\, \mathrm{Gyr} \leq \mathrm{t_{acc}} \leq 12\, \mathrm{Gyr}$. We performed a 2D Kolmogorov--Smirnov test to quantify how well each of the models matched to the observed distribution of candidate stars. This showed that models with very short evolution times (t$_{\mathrm{acc}}$ < 3 Gyr ago) are not likely to be able to explain the distribution of tidally disrupted material from $\ocen$. From this, we can place a lower limit on the accretion time of t$_{\mathrm{acc}}$ > 7 Gyr ago if we assume that $\ocen$ had an initial mass similar to its present day mass of M = $4 \times 10^6\, \mathrm{M}_\odot$, but the lower limit goes down to t$_{\mathrm{acc}}$ > 4 Gyr ago if we allow the initial mass to be larger. This is consistent with previous works which have used dynamical models to reproduce the present day orbital properties, mass, and chemical enrichment of $\ocen$ by modelling it as the remnant core of a dwarf Galaxy.

Given the scope of the current analysis, it is not possible to constrain the accretion scenario of $\ocen$ further than this. However, we emphasize that it is remarkable that publicly available data, analysis techniques, and models are now of sufficient quality to place some constraints on the accretion scenario for a single cluster using real data. In the near future, comparisons of this type with larger data-sets and improved models will be even more powerful, and will undoubtedly be a valuable tool to further unravel the accretion history of our Galaxy.

\section*{Acknowledgements}
The authors would like to thank the anonymous referee for their very useful comments which have greatly improved the quality of the paper. They would also like to thank Sven Buder (ANU), Luke Bouma (Caltech), Terese Hansen (SU), and Guillaume Thomas (IAC) for useful discussions which helped to improve the paper. KY, KL, and IK thank the European Research Council (ERC) for providing funds under the European Union's Horizon 2020 research and innovation programme (grant agreement number 852977). KY would like to thank F\aa gel\"angens catering and the Albanova cafeteria and staff, for providing delicious lunches and the daily energy required to conduct this research. This work has made use of the typesetting software \texttt{overleaf}~\footnote{\url{https://www.overleaf.com/}}, the plotting and table handling environment
\textsc{topcat} \citep{Taylor2005}, and extensively used the
\textsc{python} programming language \citep{python} for the analysis, including the following packages:
\textsc{matplotlib} \citep{Hunter2007}, \textsc{scipy} \citep{scipy}, \textsc{numpy}
\citep{Harris2020}, \textsc{pandas}
\citep{Mckinney2010}, \textsc{astropy} \citep{astropy2022, astropy2018, astropy2013} and the standard library python packages \textsc{math}¸ and \textsc{pickle} \citep{van1995python}.

\section*{Data Availability}

 The GALAH DR3 data used in this work is publicly available for download at \url{https://www.galah-survey.org/dr3/the_catalogues/}, and the list of $\ocen$ candidates is provided at the end of the paper as Table \ref{table:candidates} and online as part of the supplementary material to this paper.



\bibliographystyle{mnras}
\bibliography{galah_ocen} 




\appendix

\section{Additional Plots}
\label{sec:appendix_A}

\begin{figure*}
	\includegraphics[width=\textwidth]{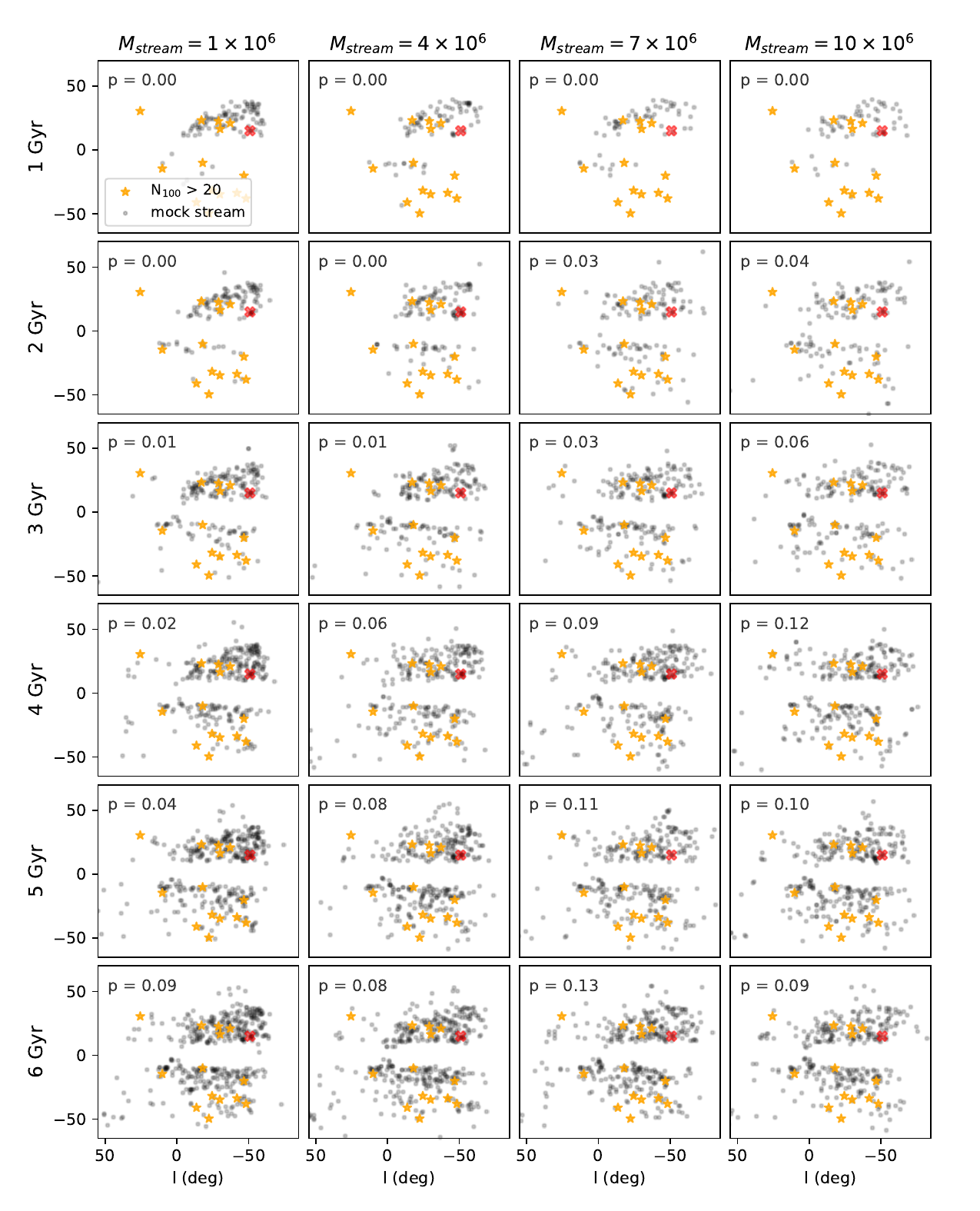}
    \caption{Grid of mock stream simulations for different initial masses and accretion times, showing the candidate stars with N$_{100} > 20$ as orange star symbols, and the masked mock stellar stream stars as black dots. The $p$-value for the KS statistic of the mock stream and candidate distribution is shown in the top left corner of each panel.}
    \label{fig:mock_stream_grid_1_6}
\end{figure*}

\begin{figure*}
	\includegraphics[width=\textwidth]{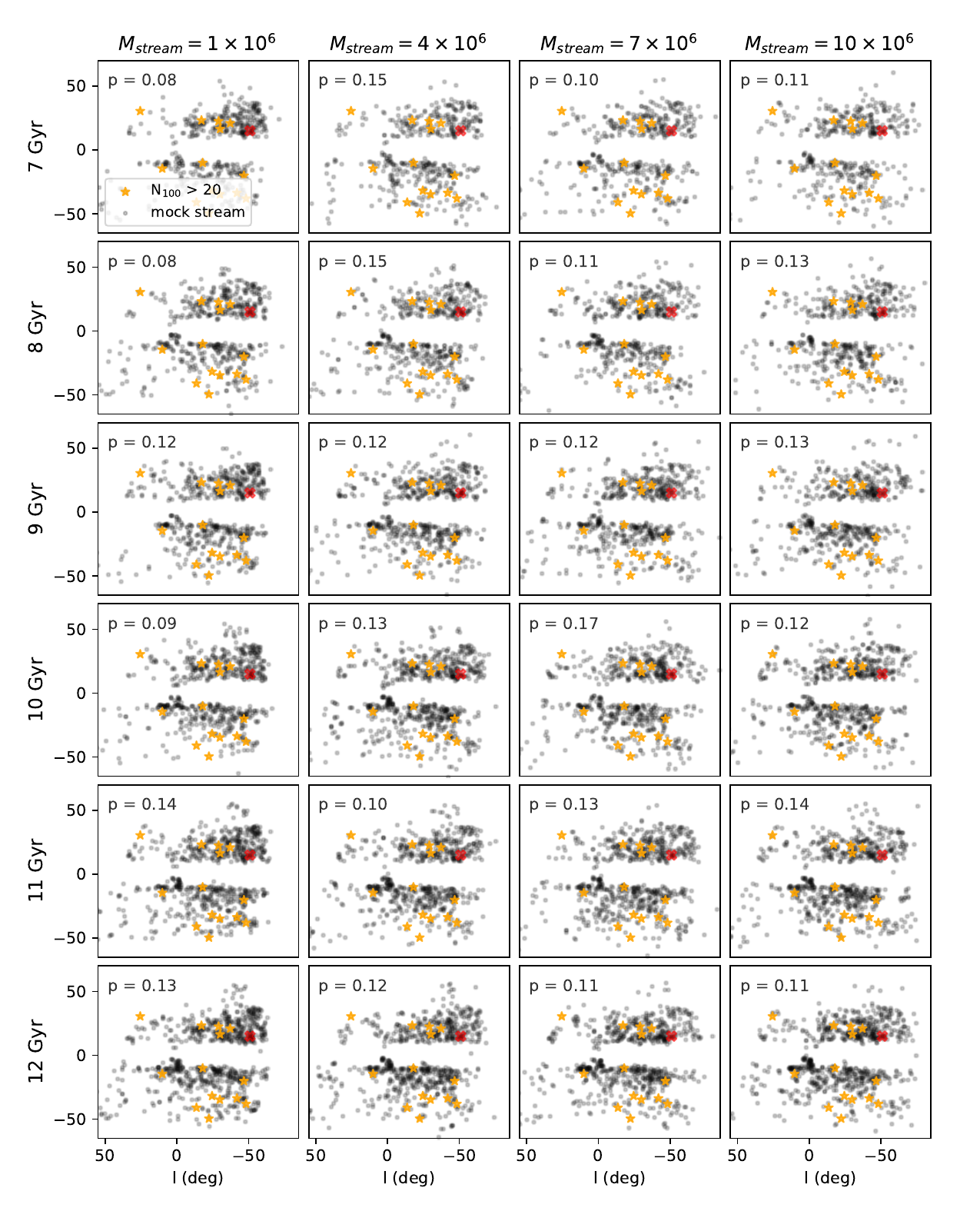}
    \caption{Continuation of Figure \ref{fig:mock_stream_grid_1_6}.}
    \label{fig:mock_stream_grid_7_12}
\end{figure*}



\begin{figure*}
	\includegraphics[width=\textwidth]{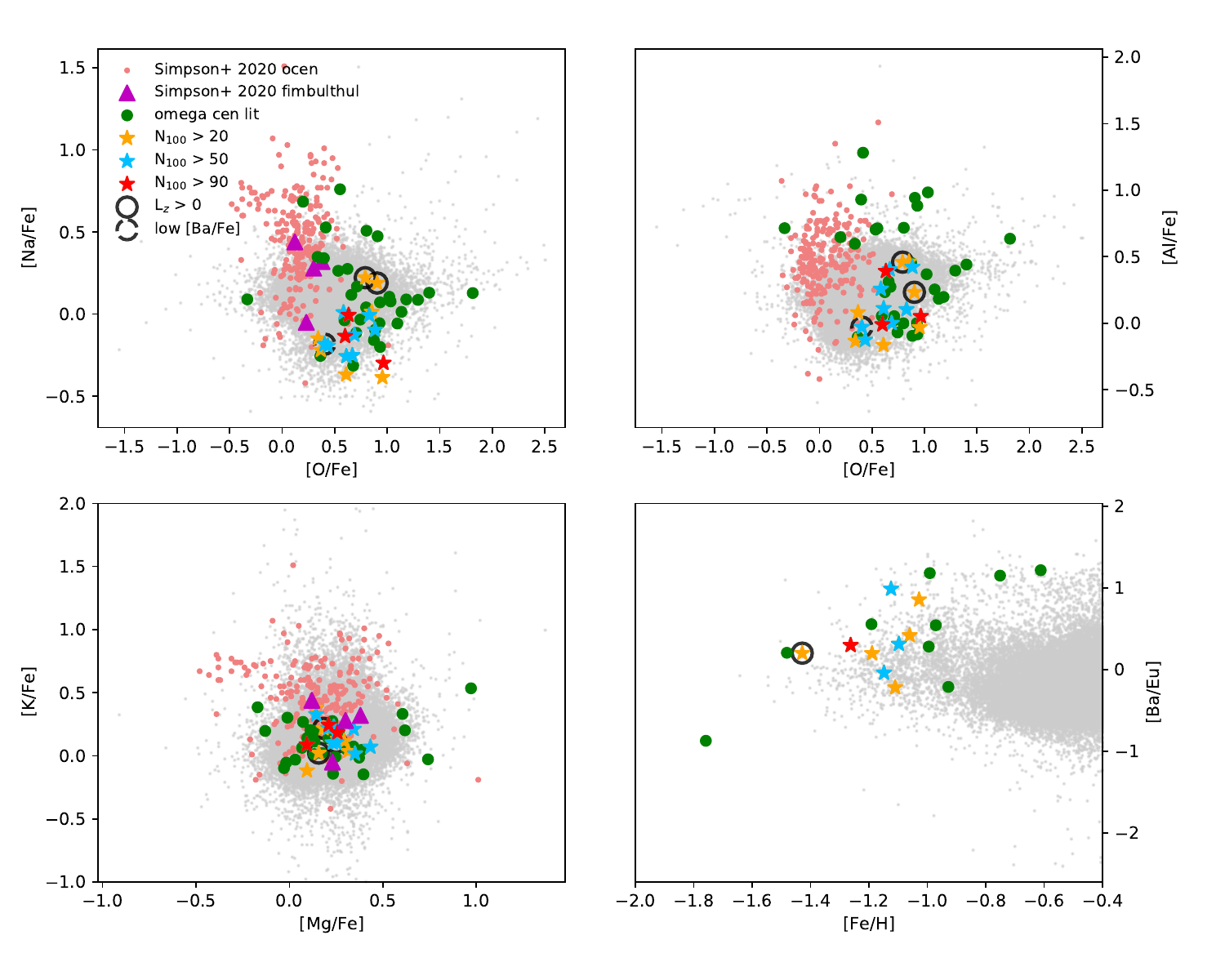}
    \caption{Anticorrelations for the $\ocen$ literature stars and t-SNE selected candidates. The bottom right panel shows the [Ba/Eu] abundances for the subset of stars which have measurements of these elements.}
    \label{fig:anti_corr_diagram}
\end{figure*}

\begin{figure*}
	\includegraphics[width=\textwidth]{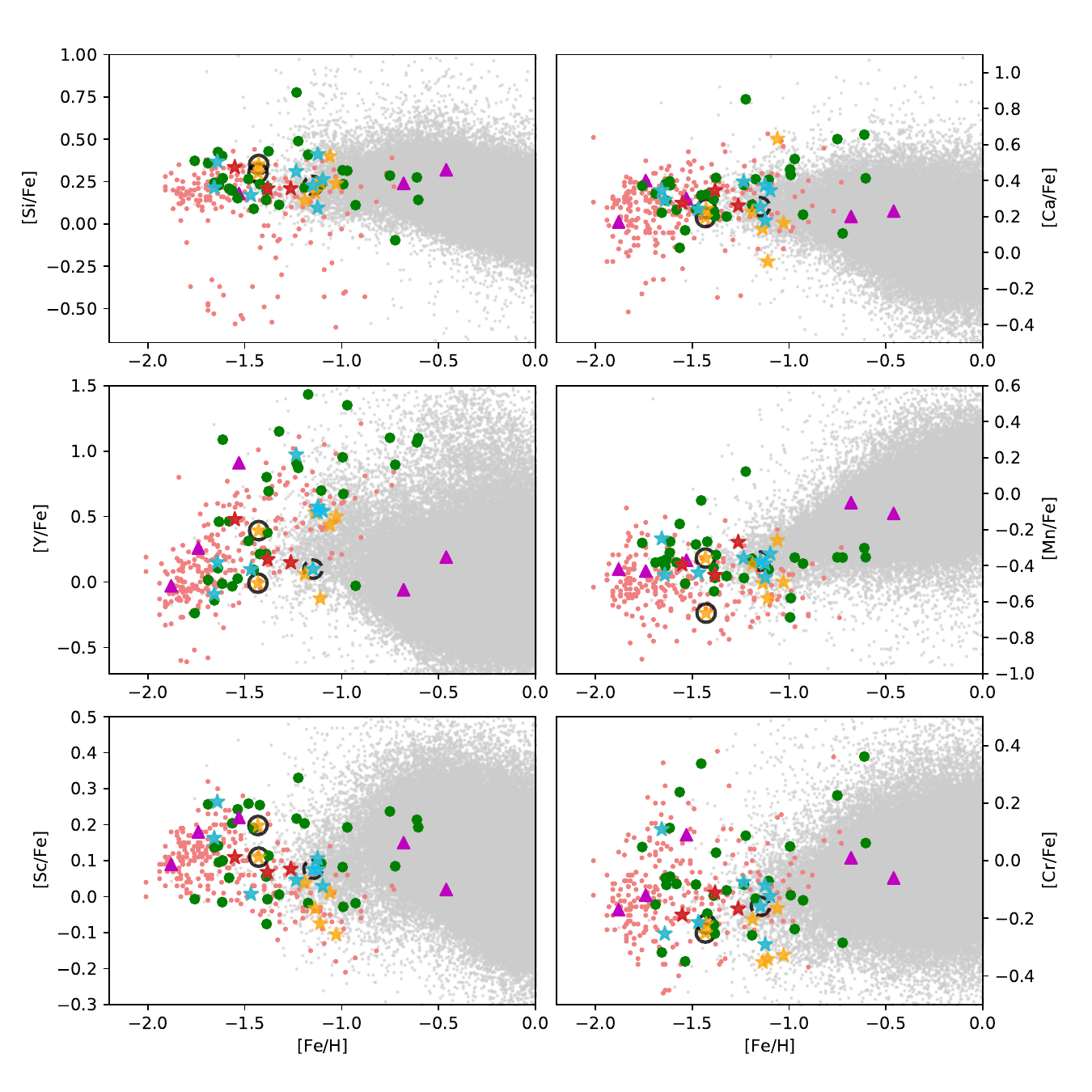}
    \caption{Chemistry of $\ocen$ candidates and confirmed cluster members. Symbols and colours are the same as in Figure \ref{fig:ocen_chemistry} and Figure \ref{fig:anti_corr_diagram}. Orange, blue and red star symbols are candidates with N$_{100} \geq$ 20, 50 and 90, respectively. Grey points are a sub-sample of the GALAH data applied and represent the bulk MW halo and disc populations. Pink points and purple triangles are the $\ocen$ literature sample and chemically selected Fimbulthul $\ocen$ candidates from \citet{Simpson2020}, respectively. Green points are $\ocen$ literature members from the GALAH sample in this work.}
    \label{fig:appendix_chemistry}
\end{figure*}

\begin{figure*}
	\includegraphics[width=\textwidth]{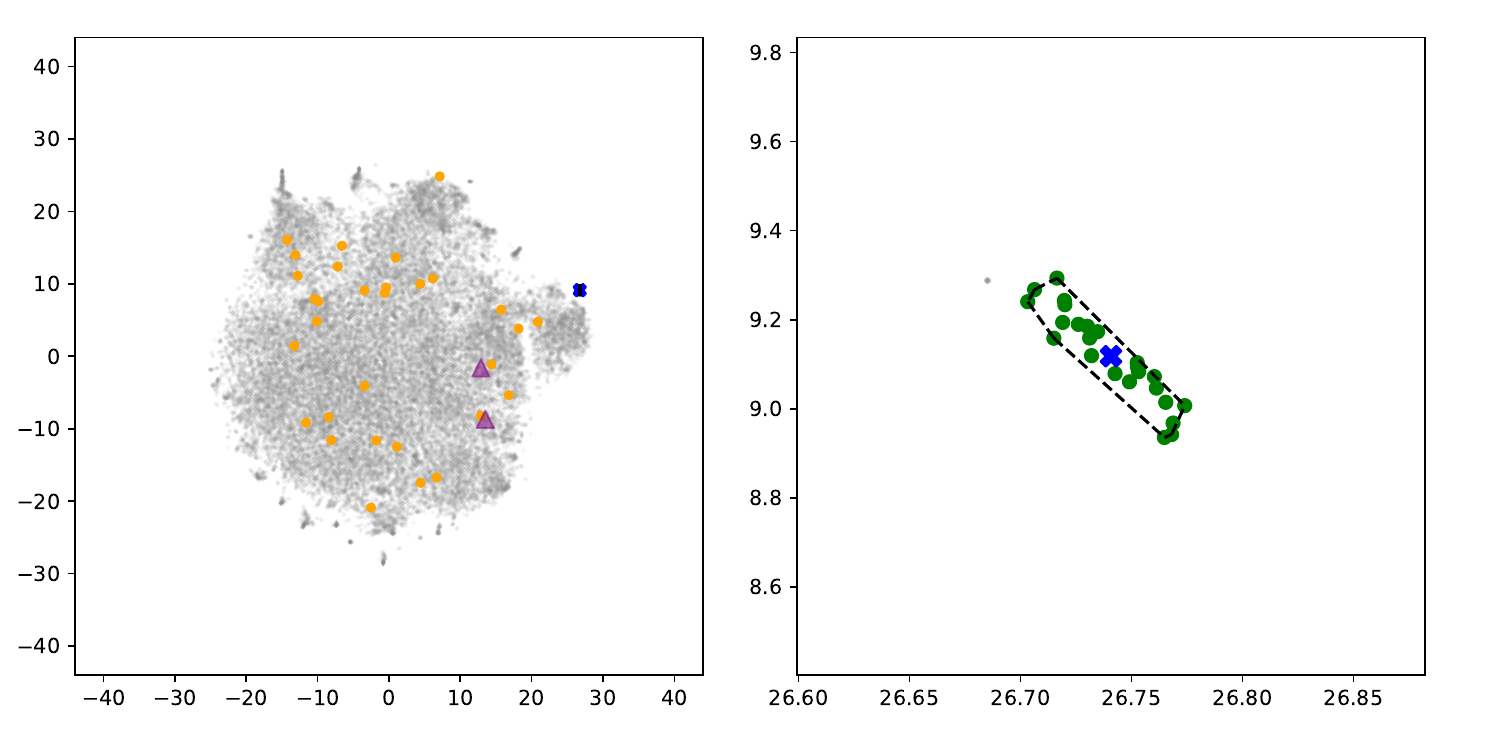}
    \caption{Latent space projection for the test using randomized, shuffled data as an input to the t-SNE. Orange points are shuffled $\ocen$ literature members, and green points are literature members with the proper parameters. Purple triangles are the \citet{Simpson2020} stars.}
    \label{fig:tsne_omega_cen_zoom_shuffle}
\end{figure*}

\begin{figure}
	\includegraphics[width=\columnwidth]{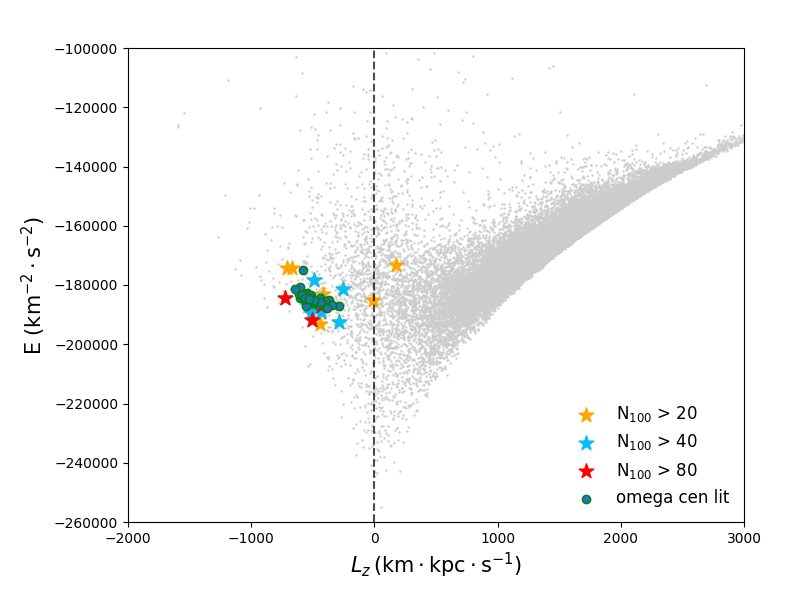}
    \caption{Energy and angular momentum space.}
    \label{fig:E_Lz_diagram}
\end{figure}

\begin{figure}
	\includegraphics[width=\columnwidth]{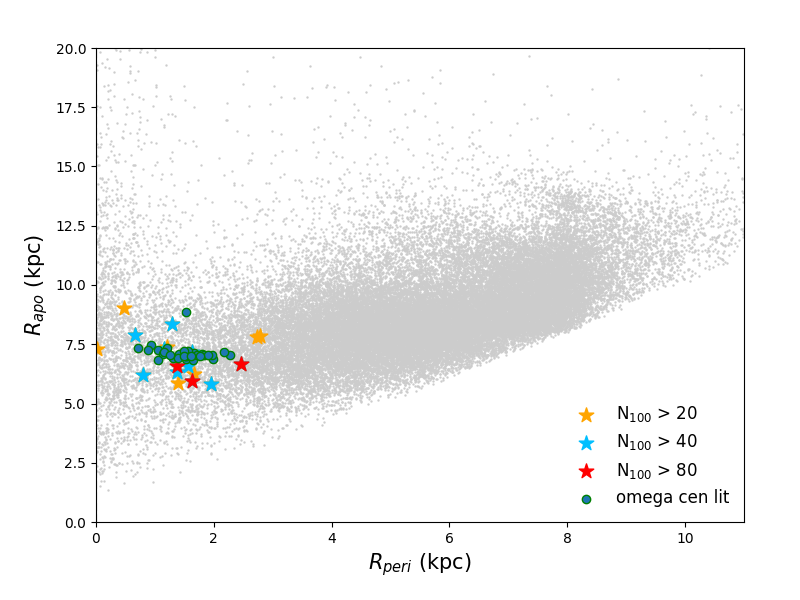}
    \caption{Pericentre radius vs apocenter radius for the selected $\ocen$ candidates.}
    \label{fig:rperi_rapo_plot}
\end{figure}

\begin{figure}
	\includegraphics[width=\columnwidth]{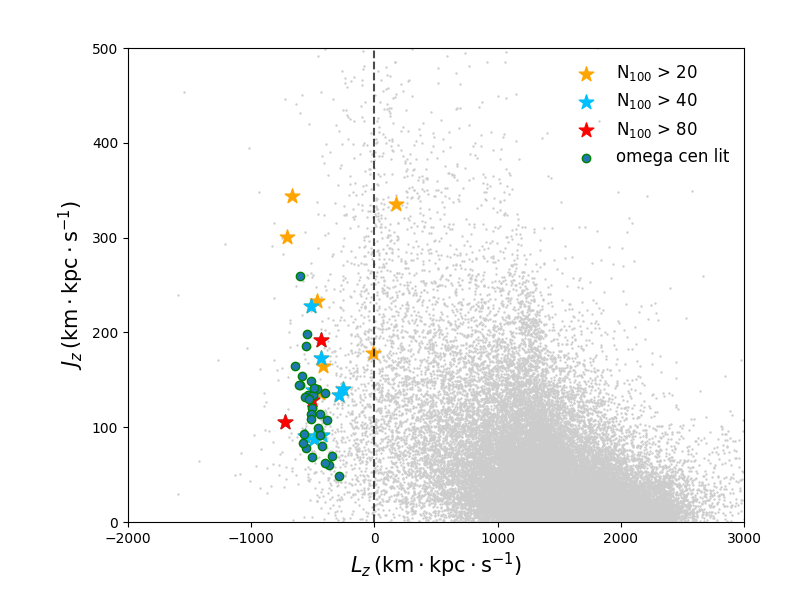}
    \caption{The z-component of the angular momentum vs the vertical action $J_z$.}
    \label{fig:Lz_Jz_diagram}
\end{figure}

\begin{figure}
	\includegraphics[width=\columnwidth]{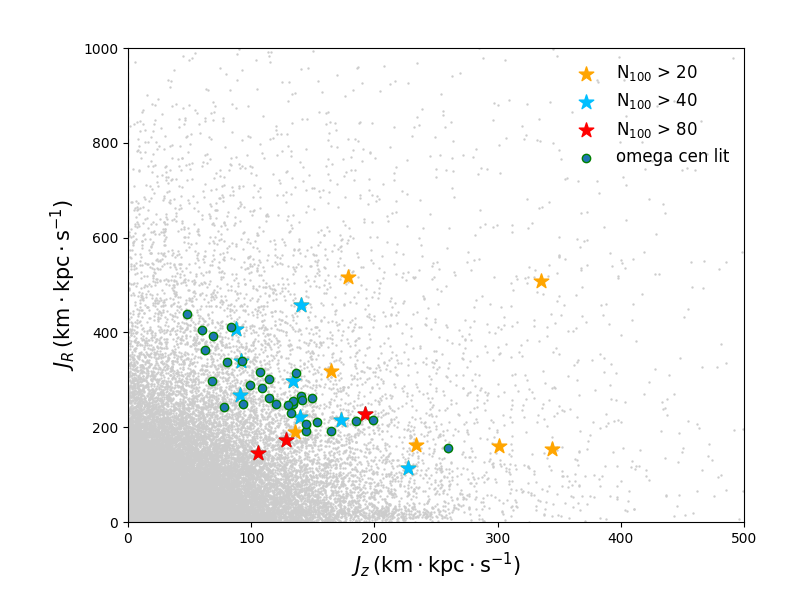}
    \caption{The vertical vs radial action components.}
    \label{fig:Jz_Jr_diagram}
\end{figure}

\begin{figure}
	\includegraphics[width=\columnwidth]{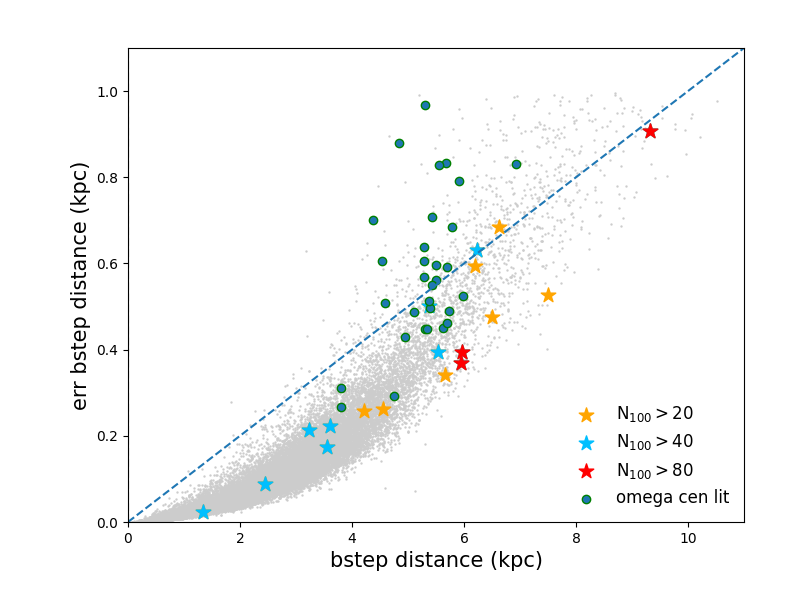}
    \caption{bstep distance error vs bstep distance. The blue dashed line shows a relative distance error of 10\%.}
    \label{fig:dist_vs_dist_err}
\end{figure}

\begin{figure}
	\includegraphics[width=\columnwidth]{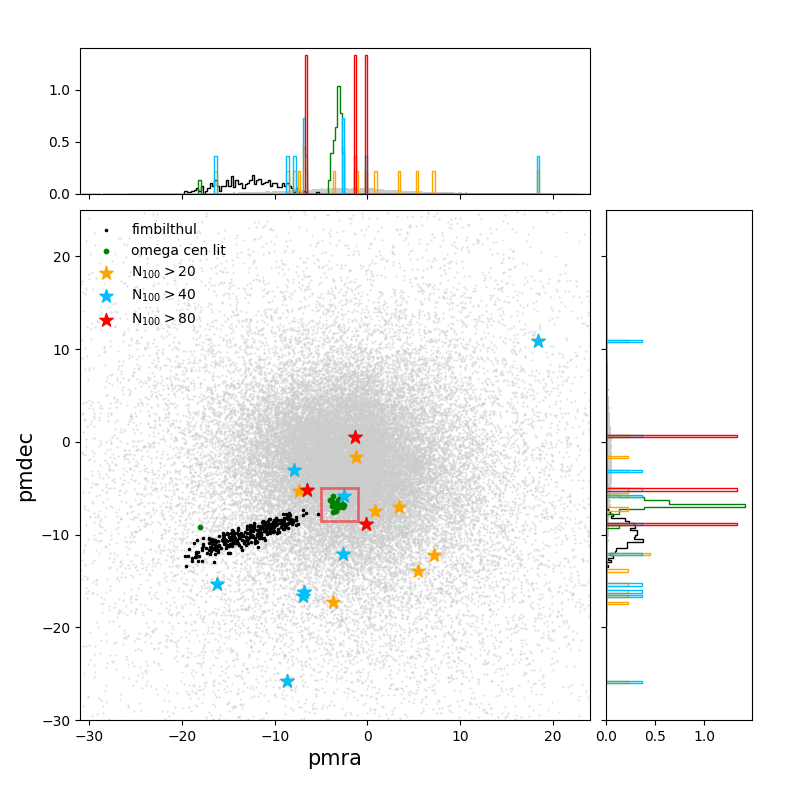}
    \caption{Proper motions of $\ocen$ literature stars, candidates, and the Fimbulthul stream. Points are coloured as in Figure \ref{fig:hist_omega_cen}. The red square outlines the region defined in \citet{Soltis2021} to contain $\ocen$ stars.}
    \label{fig:pm_diagram}
\end{figure}

\begin{figure*}
	\includegraphics[width=\textwidth]{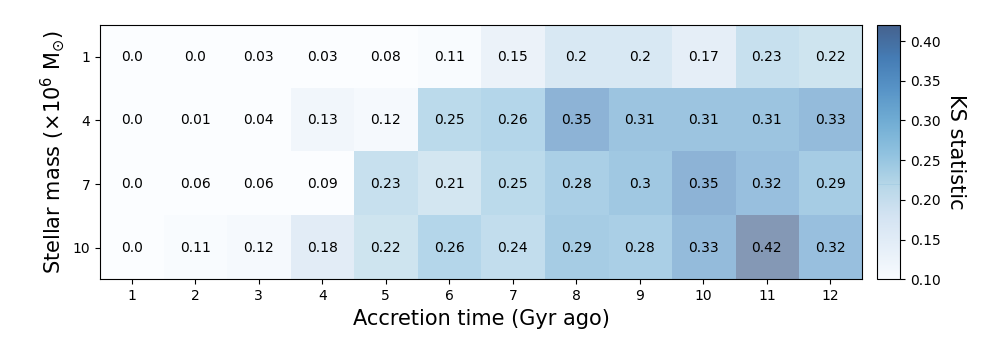}
    \caption{Grid of computed $p$-values from the 2D KS test for each combination of mass and evolution time of the mock stellar streams. Squares are coloured based on the $p$-value magnitude, and each square corresponds to the sky projections of $\ocen$ candidates with N$_{100} \geq$ 40 and mock stellar stream stars shown in Figures \ref{fig:mock_stream_grid_1_6} and \ref{fig:mock_stream_grid_7_12}.}
    \label{fig:app_KS_grid_40}
\end{figure*}

\begin{figure*}
	\includegraphics[width=\textwidth]{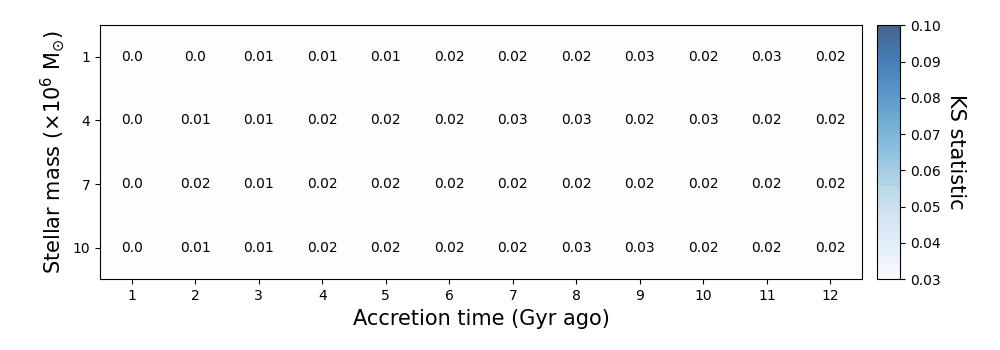}
    \caption{Grid of computed $p$-values from the 2D KS test for each combination of mass and evolution time of the mock stellar streams. Squares are coloured based on the $p$-value magnitude, and each square corresponds to the sky projections of $\ocen$ candidates with N$_{100} \geq$ 80 and mock stellar stream stars shown in Figures \ref{fig:mock_stream_grid_1_6} and \ref{fig:mock_stream_grid_7_12}.}
    \label{fig:app_KS_grid_80}
\end{figure*}

\begin{table*}
\centering
\caption{List of the 18 stars identified as \ocen\, candidates with N$_{100} \geq 20$. The bottom four rows separated by the horizontal line are the stars that were excluded from the comparison to the models based on low [Ba/Fe], L$_z \sim$ 0 or l < $-75$. This is an abbreviated version of the table, the online version also includes extra columns containing all of the abundances and orbital parameters used as input into the t-SNE.}
\label{table:candidates}
\begin{tabular}{|c|c|c|c|c|c|}
\hline
\textbf{Gaia DR3 source\_id} & \textbf{GALAH sobject\_id} & \textbf{RA}        & \textbf{Dec}  & ...      & \textbf{N$_{100}$} \\ \hline \hline
5777364572336080128            & 160331005301324             & 248.5494    & -78.2882     & ... & 93              \\
5946693318708044928            & 160529004201176             & 265.6166    & -49.4927     & ... & 90              \\
6369413689398828672            & 170531005301140             & 313.0397    & -75.4181     & ... & 81              \\
6237627600084044928            & 170603005101063             & 234.0740     & -26.5323     & ... & 77              \\
6204885586919280512            & 160425002501393             & 224.1509    & -33.4428     & ... & 67              \\
6476650325469066496            & 150606005901120             & 315.6486    & -51.9246     & ... & 55              \\
4441953003995816576            & 170407005201394             & 252.4017    & +07.6392       & ... & 44              \\
6763936500315488896            & 170531004801357             & 286.2713    & -26.6250      & ... & 42              \\
6425983425210365312            & 140809003101174             & 307.6772    & -65.6432     & ... & 42              \\
6508880485908907008            & 140805003601361             & 332.4888    & -55.2541     & ... & 34              \\
6442803577995356416            & 170515006101340             & 300.1980     & -61.4141     & ... & 34              \\
6354388244529642624            & 150705005901285             & 340.9531    & -76.8401     & ... & 31              \\
6118156899995455872            & 170602003701035             & 216.6806    & -38.2127     & ... & 30              \\
6197674401249367808            & 160524004201135             & 226.9522    & -39.2335     & ... & 22              \\ \hline
6107153807960386432            & 160522003601276             & 207.3365    & -46.1516     & ... & 68              \\
5291442147248954240            & 171207004001293             & 123.6222    & -58.3630      & ... & 56              \\ 
6349977828512542976            & 170710002701309             & 316.3541    & -79.1078     & ... & 25              \\ 
6577997436099535232            & 161013002101151             & 323.7610     & -42.3535     & ... & 23              \\ \hline
\end{tabular}
\end{table*}


\bsp	
\label{lastpage}
\end{document}